\documentclass[onecolumn,aps,floatfix,10pt,notitlepage,groupedaddress]{revtex4-2}
\pdfoutput=1
\usepackage{hyperref}
\hypersetup{colorlinks, allcolors={blue}}
\usepackage{xspace}
\usepackage{dsfont}
\usepackage{amsmath}
\usepackage{amsfonts}
\usepackage{fourier}
\usepackage{cancel}

\usepackage{graphicx}
\newcommand{\ket}[1]{\ensuremath{|#1\rangle}\xspace}
\newcommand{\bra}[1]{\ensuremath{\langle #1|}\xspace}
\newcommand{\braket}[2] {\ensuremath{\langle #1|#2\rangle}\xspace}

\renewcommand{\vec}[1]{{\mathbf{#1}}}

\newcommand{\BB}{\vec{B}}
\renewcommand{\AA}{\vec{A}}

\newcommand{\ggamma}{\boldsymbol{\gamma}}
\newcommand{\aalpha}{\boldsymbol{\alpha}}
\newcommand{\bbeta}{\boldsymbol{\beta}}
\newcommand{\rr}{\vec{r}}
\newcommand{\RR}{\vec{R}}
\newcommand{\TT}{\vec{T}}
\renewcommand{\tt}{\vec{t}}

\newcommand{\dd}{\vec{d}}
\newcommand{\bb}{\vec{b}}
\newcommand{\pp}{\vec{p}}

\newcommand{\ppi}{\boldsymbol{\pi}}

\newcommand{\hh}{\vec{h}}

\newcommand{\xx}{\hat{\vec{x}}}

\newcommand{\zz}{\hat{\vec{z}}}

\newcommand{\Frac}{\displaystyle\frac}
\newcommand{\id}{\mathds{1}}

\newcommand{\fref} [1]{Fig.~\ref{#1}}
\newcommand{\Fref} [1]{Figure~\ref{#1}}
\newcommand{\ffref}[1]{Figs.~\ref{#1}}

\newcommand{\sref} [1]{Sec.~\ref{#1}}
\newcommand{\Sref} [1]{Section~\ref{#1}}

\newcommand{\eref} [1]{Eq.~\eqref{#1}}
\newcommand{\Eref} [1]{Equation~\eqref{#1}}

\newcommand{\cref} [1]{ref.~[\onlinecite{#1}]}
\newcommand{\Cref} [1]{Reference~[\onlinecite{#1}]}

\newcommand{\aref} [1]{Appendix~\ref{#1}}

\newcommand{\aaref}[1]{Appendices~\ref{#1}}

\begin{document}

\title{Convenient Peierls phase choice for periodic atomistic systems under magnetic field}

\author{Alessandro Cresti}
\affiliation{Univ. Grenoble Alpes, Univ. Savoie Mont Blanc, CNRS, Grenoble INP, IMEP-LAHC, 38000 Grenoble, France}

\begin{abstract}
Hamiltonian models based on a localized basis set are widely used in condensed matter physics, as, for example, for the calculation of electronic structures or transport properties. 
The presence of a weak and homogeneous magnetic field can be taken into account through Peierls phase factors on the hopping Hamiltonian elements. 
Here, we propose simple and convenient recipes to properly determine such Peierls phase factors for {\it quasi}-one-dimensional systems that are periodic or with periodic subcomponents (as in Hall bars, for example), and periodic two-dimensional systems. 
We also show some examples of applications of the formulas, and more specifically concerning the electronic structure of carbon nanotubes in magnetic fields, the integer quantum Hall effect in six-terminal bars obtained in two-dimensional electron gases, and the electronic structure of bumped graphene superlattices in the presence of a magnetic field.
\\[5mm]
\noindent \emph{This is the self-archived version of the paper:} \ \ \ 	A. Cresti, Physical Review B {\bf 103}, 045402 (2021)
\\
\noindent doi: \href{\doibase 10.1103/PhysRevB.103.045402}{10.1103/PhysRevB.103.045402}
\end{abstract}

\maketitle

\section{Introduction}
Depending on the specific nature of the problem, numerical simulations in condensed matter physics can strongly benefit from the use of localized basis sets.  
In fact, the corresponding Hamiltonian is \mbox{tight-binding} like and with coupling generally limited to spatially neighbor states. 
The subsequent sparsity of the Hamiltonian matrix allows a convenient treatment of, for example, electron transport problems. 
Typically, such Hamiltonians result from the use of \mbox{semi-empirical} \mbox{tight-binding} models~\cite{Goringe1997}, from \mbox{density-functional} simulations based on localized basis sets, from the calculation of maximally localized Wannier functions~\cite{Marzari2012} for density-functional results based on plane waves, or from the direct discretization of continuous (effective mass or $k\cdot p$) Hamiltonians.

Another convenient aspect of localized basis sets is the possibility to easily take into account the presence of an external magnetic field through the \mbox{so-called} Peierls phase factors~\cite{Peierls1933}, which multiply the corresponding hopping elements in the Hamiltonian. 
These phases depend on the gauge adopted to describe the magnetic field. 
However, their circulation along a closed path is invariant and proportional to the flux of the magnetic field through the circuit itself, thus suggesting the expected invariance of the related physics observables. 
When the system or some of its parts (such as contacts and probes in a Hall bar) are periodic, a generic gauge choice will lead to a Hamiltonian that is not necessarily invariant under spatial translations. 
This prevents the applicability of convenient numerical techniques based on periodicity, such as the use of the Bloch theorem for the Hamiltonian diagonalization, or the \mbox{Sancho-Rubio} renormalization algorithm~\cite{Sancho1985} for the calculation of the contact \mbox{self-energies}. 
We can avoid these problems by a proper choice of the gauge for the magnetic field. 
In the case of {\it quasi}-one-dimensional ({\it quasi}-1D) systems, our idea is analogous to what is illustrated in \cref{Baranger1989} for continuous Hamiltonians, and already implemented in available transport codes such as Kwant~\cite{Groth2014}. 
However, our derivation moves directly from the expression of the Peierls phase factors.  
A similar approach will be illustrated for the case of two-dimensional (2D) periodic systems. 
In this case, the procedure can only be adopted for values of the magnetic fields such that the magnetic flux through the primitive cell is a multiple of the flux quantum. 
In the literature~\cite{Trellakis2003,Nemec2007,Hasegawa2013}, alternative approaches consider the use of periodic gauges specifically adapted to the geometry of the cell, or more general and sophisticated periodic gauges based on the use of singular magnetic field flux vortices. 
Our methodology has the advantage of providing a simple and explicit expression of the Peierls phase factors for arbitrarily oriented magnetic fields, and of requiring the definition of the gauge only on the discrete lattice points. 
The main result of this paper is to provide the reader with explicit and simple recipes for the Peierls phase factors, which are given in \eref{eq:peierls_phase_1d} and \eref{eq:recipe_2d_general}. 

The paper is organized as follows. 
\Sref{sec:peierls} introduces the Hamiltonian and the concept of Peierls phase. 
In \sref{sec:explicit_peierls}, we obtain the recipes to determine the Peierls phase factors for 1D and 2D systems. 
In \sref{sec:examples}, we show some examples of application of the different formulas by considering the cases of carbon nanotubes, multiterminal Hall bars and bumped graphene superlattices. 
Finally, \sref{sec:conclusions} concludes. 
Some details of the Peierls phase mathematical derivation, of the periodic gauge for 1D systems and of the relation between periodic Peierls phases for 1D and 2D systems are reported in \aaref{app:peierls}, \ref{app_vector1D}, and \ref{app:2to1}, respectively.

\section{The Peierls phase} \label{sec:peierls}
\begin{figure}[!tb] 
	\centering
		\includegraphics{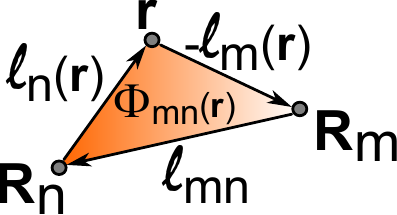}
		\caption{\label{fig_paths} 
		Linear paths $\ell_n(\rr)$ between the points $\RR_n$ and $\rr$, $\ell_m(\rr)$ between the points $\RR_m$ and $\rr$, and $\ell_{mn}$ between the points $\RR_m$ and $\RR_n$. 
		The closed path passing through the points $\RR_n$, $\RR_m$ and $\rr$, and composed of the three paths $\ell_{mn}(\rr)$, $\ell_n(\rr)$ and $-\ell_m(\rr)$, delimits the shaded area corresponding to the magnetic flux $\Phi_{mn}(\rr)$.}
\end{figure}

We preliminarily define the basis set and the Hamiltonian in the absence of the field. 
Let us consider a basis set of localized states $\{ \ket{\phi_m} \}$ with wave functions $\{ \braket{\rr}{\phi_m}=\phi_m(\rr) \}$, centered at the positions $\{\RR_m\}$, where $m$ is an integer index, and with overlap matrix $S_{mn}\equiv \braket{\phi_m}{\phi_n}$. 
For example, $\ket{\phi_m}$ could correspond to an atomic-like orbital of the atom at position $\RR_m$. 
Given a generic \mbox{one-body} Hamiltonian $\hat{H}=H(\hat{\rr},\hat{\pp})$, which is function of the position and momentum operators $\hat{\rr}$ and $\hat{\pp}$, the corresponding matrix elements on the considered basis set are
\begin{equation}
   H_{mn} \ \equiv \ \braket{\phi_m|\hat{H}}{\phi_n} \ = \ \int dr^3 \ \phi_m^*(\rr) \ H(\rr,-i\hbar\nabla) \ \phi_n(\rr) \ ,
\end{equation}
where we considered the real-space representation of the momentum operator $\hat{\pp}\rightarrow -i\hbar\nabla$. 
Other degrees of freedom, such as the spin, can be straightforwardly included without entailing any modification of the procedure illustrated in this section. 
In the presence of a static magnetic field $\BB(\rr)$ described by a vector potential $\AA(\rr)$ through the relation $\BB(\rr)=\nabla\times\AA(\rr)$, the Hamiltonian transforms according to the principle of minimal coupling~\cite{Jackson1998} as
\begin{equation}
   \hat{H}_\AA \ = \ H(\hat{\rr},\hat{\ppi}) \ \ \ \ \ {\rm with} \ \ \ \ \ \hat{\ppi} \ = \ \hat{\pp} \ + \ \Frac{e}{c} \AA(\hat{\rr}) \ ,
\end{equation}
where $e>0$ is the absolute value of the electron charge and $c$ is the speed of light. 
We proceed by defining a new set of localized states $\{ \ket{\tilde{\phi}_n} \}$ with wave functions~\cite{Peierls1933,Luttinger1951}
\begin{equation} \label{eq:new_set}
   \tilde{\phi}_n(\rr) \ = \ \braket{\rr}{\tilde{\phi}_n} \ \equiv \ \exp\left[-i\Frac{e}{\hbar c}\int_{\ell_n(\rr)} \!\!\!\!\!\!\!\! d\rr'\cdot\AA(\rr')   \right] \ \phi_n(\rr)
	  \ = \ \exp\left[-i\Frac{e}{\hbar c}G_n(\rr)\right] \ \phi_n(\rr)
		\ \ \ \ \ \ \ {\rm with} \ \ \ \ \ \ \ 
	  G_n(\rr) \ \equiv \ \int_{\ell_n(\rr)} \!\!\!\!\!\!\!\! d\rr'\cdot\AA(\rr')
	\ ,
\end{equation}
where $\ell_n(\rr)$ is the straight line path from $\RR_n$ to $\rr$, see \fref{fig_paths}, which we parametrize as $\ell_n(\rr,\lambda) = \RR_n  +  (\rr-\RR_n)  \lambda$ in terms of $\lambda\in[0,1]$. 
As a consequence, the line integral $G(\rr)$ becomes
\begin{equation} \label{eq:Gn}
		G_n(\rr) \ = \ \int_0^1 \ d\lambda \ (\rr-\RR_n)\cdot\AA\left(\RR_n + (\rr-\RR_n) \ \lambda\right) \ .
\end{equation}
If we apply the momentum operator $\hat{\ppi}$, which corresponds to $-i\hbar\nabla+e/c \AA(\rr)$ in the real-space representation, we obtain
\begin{equation} \label{eq:momentum_applied_preliminary}
    \braket{\rr|\hat{\ppi}}{\tilde{\phi}_n} 
		\ = \ \left[-i\hbar\nabla \ + \ \Frac{e}{c} \AA(\rr) \right] \exp\left[-i\Frac{e}{\hbar c}G_n(\rr) \right] \ \phi_n(\rr) 
		\ = \ \exp\left[-i\Frac{e}{\hbar c}G_n(\rr) \right] \ \left[-i\hbar\nabla \phi_n(\rr) \ - \ \Frac{e}{c} \nabla G_n(\rr) \ \phi_n(\rr)  \ + \ \Frac{e}{c} \AA(\rr) \ \phi_n(\rr) \right] \ . 		
\end{equation}
As detailed in \aref{app:peierls}, according to the derivation of \cref{Luttinger1951} we can demonstrate that, for a well-localized basis set $\{\ket{\phi_m}\}$, we have $\nabla G_n(\rr)\approx \AA(\rr)$.
As a consequence, the last two terms of \eref{eq:momentum_applied_preliminary} cancel and 
\begin{equation} \label{eq:momentum2_applied}
    \braket{\rr|\hat{\ppi}}{\tilde{\phi}_n} 
		\ = \  \exp\left[-i\Frac{e}{\hbar c}G_n(\rr) \right]  \braket{\rr|\hat{\pp}}{\phi_n} 
		\ = \  \exp\left[-i\Frac{e}{\hbar c}\int_{\ell_n(\rr)} \!\!\!\!\!\!\!\! d\rr'\cdot\AA(\rr') \right]  \braket{\rr|\hat{\pp}}{\phi_n} \ .
\end{equation}
In a similar manner, we can demonstrate that 
\begin{equation} \label{eq:momentum_applied}
    \braket{\rr|\hat{\ppi}^2}{\tilde{\phi}_n} 
		\ = \  \exp\left[-i\Frac{e}{\hbar c}G_n(\rr) \right]  \braket{\rr|\hat{\pp}^2}{\phi_n} 
		\ = \  \exp\left[-i\Frac{e}{\hbar c}\int_{\ell_n(\rr)} \!\!\!\!\!\!\!\! d\rr'\cdot\AA(\rr') \right]  \braket{\rr|\hat{\pp}^2}{\phi_n}
\end{equation}
and then 
\begin{eqnarray}
    \left[H_\AA\right]_{mn} & = & \braket{\tilde{\phi}_m|\hat{H}_\AA}{\tilde{\phi}_n} 
	                 \ = \ \int dr^3  
		  							\exp\left[ i\Frac{e}{\hbar c}\int_{\ell_m(\rr)} \!\!\!\!\!\!\!\! d\rr'\cdot\AA(\rr') \right] \ \phi_m^*(\rr) H\left(\rr,-i\hbar\nabla+\Frac{e}{c}\AA(\rr)\right)
										\exp\left[-i\Frac{e}{\hbar c}\int_{\ell_n(\rr)} \!\!\!\!\!\!\!\! d\rr'\cdot\AA(\rr') \right]   \phi_n(\rr)  \nonumber
				\\[5mm]		& = & \ \int dr^3  
		  							\exp\left[-i\Frac{e}{\hbar c}\int_{\ell_n(\rr)} \!\!\!\!\!\!\!\! d\rr'\cdot\AA(\rr')
										          +i\Frac{e}{\hbar c}\int_{\ell_m(\rr)} \!\!\!\!\!\!\!\! d\rr'\cdot\AA(\rr')  \right] \
										 \phi_m^*(\rr)H(\rr,-i\hbar\nabla) \phi_n(\rr)            \nonumber
				\\[5mm] 	& = & \ \int dr^3  
		  							\exp\left[-i\Frac{e}{\hbar c}\int_{\ell_n(\rr)-\ell_m(\rr)} \!\!\!\!\!\!\!\! d\rr'\cdot \AA(\rr') \right] \
										 \phi_m^*(\rr)H(\rr,-i\hbar\nabla) \phi_n(\rr)            \nonumber
				\\[5mm] 	& = & \exp\left[-i\Frac{e}{\hbar c}\int_{-\ell_{mn}}\!\!\!\!\!\!\!\! d\rr'\cdot \AA(\rr') \right] \
									      \int dr^3  
		  							    \exp\left[-i\Frac{e}{\hbar c}\int_{\ell_n(\rr)-\ell_m(\rr)+\ell_{mn}} \!\!\!\!\!\!\!\!\!\!\!\!\!\!\!\!\!\!\!\!\!\!\!\! d\rr'\cdot\AA(\rr') \right] \
										 	 \phi_m^*(\rr)H(\rr,-i\hbar\nabla) \phi_n(\rr)  \ ,      \nonumber
				\\[5mm] 	& = & \exp\left[-i\Frac{e}{\hbar c}\int_{-\ell_{mn}}\!\!\!\!\!\!\!\! d\rr'\cdot \AA(\rr') \right] \
									      \int dr^3  
		  							    \exp\left[-i\Frac{e}{\hbar c}\Phi_{mn}(\rr) \right] \
										 	 \phi_m^*(\rr)H(\rr,-i\hbar\nabla) \phi_n(\rr)  \ , \label{eq:ham_phase}
\end{eqnarray}
where we made use of \eref{eq:momentum_applied} to obtain the second line, we introduced the straight line path $\ell_{mn}$ from $\RR_m$ to $\RR_n$, as indicated in \fref{fig_paths}, and defined 
\begin{equation} \label{eq:flux_integral}
  \Phi_{mn}(\rr) \ \equiv \ \int_{\ell_n(\rr)-\ell_m(\rr)+\ell_{mn}} \!\!\!\!\!\!\!\!\!\!\!\!\!\!\!\!\!\!\!\!\!\!\!\! d\rr'\cdot\AA(\rr') \ ,
\end{equation}
which is none other than the gauge-invariant flux of the magnetic field through the area enclosed by the circuit $\ell_n(\rr)-\ell_m(\rr)+\ell_{mn}$, corresponding to the shaded region in \fref{fig_paths}. 
Due to the localized nature of the basis, the second integrand in \eref{eq:ham_phase} is expected to be non-vanishing only when $\rr$ is around $\RR_m$ or $\RR_n$. 
When $\rr$ is in this region, the choice of $\ell_{mn}$ as a linear path reduces the surface enclosed by the circuit $\ell_n(\rr)-\ell_m(\rr)+\ell_{mn}$, and then the magnetic flux $\Phi_{mn}(\rr)$. 
Moreover, if the magnetic field varies smoothly on the scale of the atomic positions, the flux changes its sign when $\rr$ moves from one side to the other of the line $\ell_{mn}$. 
As a consequence, we can approximate $\Phi_{mn}(\rr)\approx 0$ in \eref{eq:ham_phase}, and obtain
\begin{equation}
    \left[H_\AA\right]_{mn} \approx \exp\left[-i\Frac{e}{\hbar c}\int_{-\ell_{mn}}\!\!\!\!\!\!\!\! d\rr'\cdot \AA(\rr') \right] \
									      \int dr^3  \phi_m^*(\rr)H(\rr,-i\hbar\nabla) \phi_n(\rr)  
										\ = \ \exp\left[-i\Frac{e}{\hbar c}\int_{-\ell_{mn}}\!\!\!\!\!\!\!\! d\rr'\cdot \AA(\rr') \right] \ H_{mn}
										\ = \ e^{i\varphi_{mn}} \ H_{mn} \ .
\end{equation}
We define the Peierls phase $\varphi_{mn}$~\cite{Peierls1933} as
\begin{equation} \label{eq:peierls_phase}
  \varphi_{mn}  \ \equiv \ \Frac{2\pi}{\Phi_0}\int_{\RR_m}^{\RR_n}\! d\rr \cdot \AA(\rr) \ ,
\end{equation}
where the integral is performed along the linear path $\ell_{mn}$ from $\RR_m$ to $\RR_n$, and $\Phi_0=hc/e\approx$4.136$\times$10$^{-15}$~Wb is the flux quantum. 
The empirical choice of a linear path for $\ell_{mn}$ is widely adopted in the literature~\cite{Graf1995}, and can be justified in a more rigorous way from gauge invariance~\cite{Boykin2001}. 
The Peierls phase is not gauge independent since
\begin{equation} \label{eq:gauge_invariance}
  \AA(\rr) \ \rightarrow \ \AA(\rr) \ + \ \nabla\chi(\rr) \ \ : \ \ \varphi_{mn} \ \rightarrow \  \varphi_{mn} \  + \ \Frac{2\pi}{\Phi_0} \left[\chi_n\ - \ \chi_m\right] \ ,
\end{equation}
where $\chi(\rr)$ is any differentiable scalar function and $\chi_m\equiv \chi(\RR_m)$. 
However, as is evident from \eref{eq:gauge_invariance}, the circulation of the Peierls phase along a closed path is \mbox{gauge-independent}, being proportional to the flux of the magnetic field through the surface enclosed by the path. 
Note that the gauge $\chi(\rr)$ does not need to vary slowly with respect to the interatomic distance, but just needs to be a continuous function. Indeed, at the position $\RR_i$, $\chi_i$ can be freely chosen, since we can always define a continuous function with given values on a discrete set of points. 

The overlap matrix $S$ changes in the same way as the Hamiltonian, as seen by substituting the operator $\hat{H}_\AA$ with the identity operator $\hat{\id}$ in the previous derivation, i.e.,
\begin{equation} \label{eq:new_overlap}
    \tilde{S}_{mn} \ \equiv \ \braket{\tilde{\phi}_m}{\tilde{\phi}_n} \approx \ e^{i\varphi_{mn}} \ S_{mn} \ .
\end{equation}
The results presented in \sref{sec:explicit_peierls} are thus valid for a general (also nonorthonormal) basis, provided \eref{eq:new_overlap} is taken into account. 
If the starting basis set is orthonormal, from \eref{eq:new_overlap} it follows that also the basis set defined by \eref{eq:new_set} is approximately orthonormal, as in the case of the examples of \sref{sec:examples}.

We remark that the literature~\cite{London1937,Pople1962} also proposes a phase choice that is an alternative to that of \eref{eq:new_set}, specifically
\begin{equation} \label{eq:new_set_alternative}
   \tilde{\phi}_n(\rr) \ = \ \braket{\rr}{\tilde{\phi}_n} \ \equiv \ \exp\left[-i\Frac{e}{\hbar c} \ \AA(\RR_n)\cdot\rr   \right] \ \phi_n(\rr) \ .
\end{equation}
While this different choice entails a different mathematical treatment compared to \eref{eq:new_set}, under the commonly adopted approximations, the resulting Peierls phase is equivalent to that of~\eref{eq:peierls_phase}.
More accurate approaches, which take into account further terms, for the Hamiltonian and the overlap matrix are available in the literature~\cite{Matsuura2016}. 

\section{Explicit Peierls phase factors for a homogeneous magnetic field} \label{sec:explicit_peierls}
Given a homogeneous magnetic field $\BB$, we can write
\begin{equation} \label{eq:peierls_general}
  \AA(\rr) \ = \ \Frac{1}{2} \ \BB\times\rr \ + \ \nabla\chi(\rr) \ .
\end{equation}
The corresponding Peierls phase $\varphi_{mn}$ is obtained by performing the integration of \eref{eq:peierls_phase} along the straight line from $\RR_m$ to $\RR_n$:
\begin{eqnarray} \label{eq:peierls_phase_general}
  \varphi_{mn}  & = & \Frac{2\pi}{\Phi_0}\left[ 
	\left( \int_{\RR_m}^{\RR_n}\! d\rr \cdot \Frac{1}{2} \ \left(\BB\times\rr\right) \right)
	\ + \ \chi(\RR_n) \ - \ \chi(\RR_m) \right] 
			\ = \ \Frac{2\pi}{\Phi_0}\left[ \Frac{1}{2} \ \BB \cdot \left( \RR_m \times \RR_n \right) \ + \ \chi(\RR_n) \ - \ \chi(\RR_m) \right] \ ,
\end{eqnarray}
where the generic gauge function $\chi$ on the discrete points $\{\RR_m\}$ is defined {\it modulo} $\Phi_0$. 
\begin{figure}[!tb]
	\centering
		\resizebox{14cm}{!}{\includegraphics{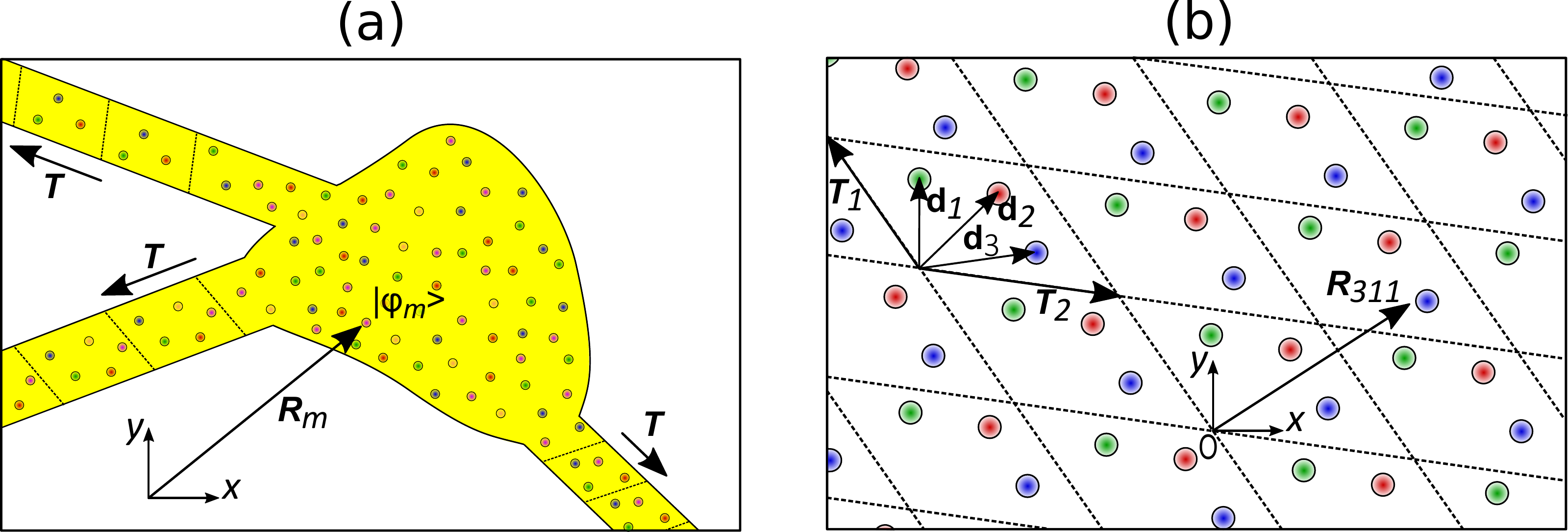}}
		\caption{ \label{fig_1d_2d_sketch}(a) System with periodic regions, for example leads, with translation vector $\TT$, which can be different for the different leads. (b) Periodic 2D system. The translation vectors are $\TT_1$ and $\TT_2$, and the basis vectors are indicated by $\{\dd_i\}$.}
\end{figure}
%

\subsection{One-dimensional periodic systems}
Let us consider a system that is invariant under translations by a vector $\TT$. 
If we require also the Hamiltonian to be invariant, the Peierls phase $\varphi_{\bar{m}\bar{n}}$ for a translated couple of states, i.e., such that $\RR_{\bar{m}}=\RR_m+\TT$ and $\RR_{\bar{n}}=\RR_n+\TT$, must be equal to $\varphi_{mn}$, {\it modulo} $2\pi$.
Therefore, according to \eref{eq:peierls_phase_general}, we have
\begin{equation} \label{eq:1d_condition}
  0  = \varphi_{\bar{m}\bar{n}}  -  \varphi_{mn}  =  
	\Frac{2\pi}{\Phi_0}\left[ \chi(\RR_n+\TT)  -  \Frac{1}{2}  \RR_n\cdot(\TT\times\BB)  - \chi(\RR_n) \right]
- \Frac{2\pi}{\Phi_0} \left[ \chi(\RR_m+\TT)  -  \Frac{1}{2}  \RR_m\cdot(\TT\times\BB)  -  \chi(\RR_m) \right] \ ,
\end{equation}
{\it modulo} $2\pi$. 
Such a condition is satisfied if
\begin{equation}  \label{eq:1d_condition2}
	\chi(\RR_m+\TT) \ = \ \Frac{1}{2} \ \RR_m\cdot(\TT\times\BB) \ + \ \chi(\RR_m) \ .
\end{equation}
From the previous formula, we deduce that $\chi(\rr)$ is at least quadratic in $\rr$. 
We guess the function  
\begin{equation}
	\chi(\rr) \ = \ \rr\cdot\aalpha \ \ \rr\cdot\bbeta \ + \ \ggamma\cdot\rr \ + \ \eta \ ,
\end{equation}
where $\aalpha$, $\bbeta$, and $\ggamma$ are vectors and $\eta$ is a scalar.
Therefore, \eref{eq:1d_condition2} becomes
\begin{equation}
	\chi(\RR_m+\TT) \ - \ \chi(\RR_m) \ = \ \TT\cdot\aalpha \ \ \RR_m\cdot\bbeta
	                              \ + \ \RR_m\cdot\aalpha \ \ \TT\cdot\bbeta
																\ + \ \TT\cdot\aalpha \ \ \TT\cdot\bbeta
																\ + \ \ggamma\cdot\TT
																\ = \ \Frac{1}{2} \ \RR_m\cdot(\TT\times\BB) \ .
\end{equation}
A simple and convenient choice to solve this equation is
\begin{equation} \label{eq:gauge1d}
		\alpha \ = \ \Frac{\TT}{2\left|\TT\right|}
		\ \ ; \ \
		\bbeta \ = \ \Frac{\TT}{\left|\TT\right|}\times\BB
		\ \ ; \ \
		\ggamma \ =  0
		\ \ ; \ \
		\eta \ = \ 0
		\ \ \ \ \ \rightarrow \ \ \ \ \ 
		\chi(\rr) \ = \ \Frac{1}{2} \ \rr\cdot\tt\ \ \rr\cdot\left( \tt\times\BB \right)
		\ \ {\rm with} \ \
		\tt \ \equiv \ \Frac{\TT}{\left|\TT\right|} \ .
\end{equation}
Note that the resulting Peierls phase is also invariant under spatial translations along the direction of the magnetic field. 
More details about the vector potential of \eref{eq:peierls_general} with the gauge choice of \eref{eq:gauge1d} are illustrated in \aref{app_vector1D}. 
In particular, it is straightforward to show that if $\tt=\xx$ and $\BB=B\ \zz$, as in the case  of a carbon nanotube oriented along the $x$~direction in a homogeneous magnetic field along the $z$~direction discussed in \sref{subsec:cnts}, then the resulting vector potential is $\AA(\rr)=- B \ y \ \xx$, which does correspond to the first Landau gauge usually adopted in the literature~\cite{Saito1998,Torres2020}. 

In case we deal with, for example, a multiterminal system with leads oriented along different directions, see \fref{fig_1d_2d_sketch}(a), we can define the gauge in each region according to the periodicity direction $\tt(\rr)$ of the region corresponding to the position $\rr$. 
If the position $\rr$ corresponds to a nonperiodic region (for example the central part of a device or a region where the periodicity of the phase is not required), we can choose any $\tt(\rr)$, for example $\tt(\rr)=\vec{0}$. 

In conclusion, the recipe to calculate the Peierls phase is
\begin{equation} \label{eq:peierls_phase_1d}
  \varphi_{mn} \ = \ \Frac{\pi}{\Phi_0} \ \BB \cdot \left\{ \RR_m \times \RR_n 
																                    \ + \ \left[\RR_n\cdot\tt(\RR_n)\right] \ \RR_n \times \tt(\RR_n) \ 
							                    	                \ - \ \left[\RR_m\cdot\tt(\RR_m)\right] \ \RR_m \times \tt(\RR_m) \
																										\right\} \ .
\end{equation}
%

\subsection{Two-dimensional periodic systems}
Periodic 2D lattices are defined by a unit cell repeated periodically with translation vectors $\TT_1$ and $\TT_2$. 
The basis set $\{\ket{\phi_{i m_1 m_2}} \}$ consists of states localized around the positions $\{ \RR_{i m_1 m_2} = \dd_i + m_1 \TT_1 + m_2 \TT_2 \}$, where $\dd_i$ indicates the position of the \mbox{$i$th} state inside the unit cell and $m_1 \TT_1 + m_2 \TT_2$ is the position of the cell with integer indices $m_1$ and $m_2$, see \fref{fig_1d_2d_sketch}(b). 
According to \eref{eq:peierls_phase_general}, the Peierls phase between the two states at positions $\RR_{im_1m_2}$ and $\RR_{jn_1n_2}$ is
\begin{equation}
  \varphi_{i,m_1,m_2;j,n_1,n_2} 
			    = \displaystyle \frac{2\pi}{\Phi_0} \ \ \left[ \frac{1}{2} \ \ \BB\cdot \left( \RR_{im_1m_2}\times\RR_{jn_1n_2} \right) 
                          \ + \ \chi(\RR_{jn_1n_2}) \ - \ \chi(\RR_{im_1m_2}) \right] \ .
\end{equation}
To ensure Hamiltonian periodicity, the Peierls phase must be invariant {\it modulo} $2\pi$ under any translation $\TT_{s_1 s_2}=s_1 \TT_1+s_2 \TT_2$, i.e., $\varphi_{i,m_1+s_1,m_2+s_2;j,n_1+s_1,n_2+s_2}=\varphi_{im_1m_2;jn_1n_2}+ 2\pi q$, for any integer numbers $s_1$ and $s_2$, where $q\in\mathbb{Z}$. 
Therefore
\begin{eqnarray} \label{eq:dphi} 2\pi q & = & \varphi_{i,m_1+s_1,m_2+s_2;j,n_1+s_1,n_2+s_2} - \varphi_{i,m_1,m_2;j,n_1,n_2}
                   \\[3mm] & = & \Frac{2\pi}{\Phi_0} \left[
											\chi_{j,n_1+s_1,n_2+s_2} - \chi_{j,n_1,n_2} - \Frac{1}{2} \RR_{j n_1 n_2} \cdot \left( \TT_{s_1 s_2} \times \BB \right) 
											- \chi_{i,m_1+s_1,m_2+s_2} + \chi_{i,m_1,m_2} + \Frac{1}{2} \RR_{i m_1 m_2} \cdot \left( \TT_{s_1 s_2} \times \BB \right) \right]
\ . \nonumber
\end{eqnarray}
Let us define the function $g(i,m_1,m_2,s_1,s_2)$ such that
\begin{eqnarray} \label{eq:chi_a}
      \chi_{i,m_1+s_1,m_2+s_2} & \equiv &  \chi_{i m_1 m_2}  + \Frac{1}{2} \RR_{i m_1 m_2} \cdot \left( \TT_{s_1 s_2} \times \BB \right) \ + \ g(i,m_1,m_2,s_1,s_2)
			\ ,
\end{eqnarray}
\textit{modulo} $\Phi_0$. 
To satisfy \eref{eq:dphi}, we must have
\begin{equation}
	g(j,n_1,n_2,s_1,s_2) \ = \ g(i,m_1,m_2,s_1,s_2) \ + \ q \Phi_0 \ \ \ \forall \ i,\ j,\ m_1,\ m_2,\ n_1,\ n_2,\ s_1,\ s_2 \ \ \ {\rm with} \ \ \ q\in\mathbb{Z} \ ,
\end{equation}
which implies, in the most general case, that $g(i,m_1,m_2,s_1,s_2)=f(s_1,s_2)$, i.e., it only depends on its last two arguments $s_1$ and $s_2$. 
By setting $s_1=m_1$, $s_2=m_2$, $m_1=0$ and $m_2=0$ in \eref{eq:chi_a}, we obtain
\begin{eqnarray} \label{eq:chi_b}
      \chi_{i,m_1,m_2} & = &  \chi_{i 0 0} \  + \ \Frac{1}{2} \BB \cdot \left( \dd_i \times \TT_{m_1 m_2}  \right) \ + \ f(m_1,m_2) \ . 
\end{eqnarray}
From \eref{eq:chi_a} and \eref{eq:chi_b}, it follows that
\begin{eqnarray} \label{eq:chi_b*}
  \chi_{i,m_1+s_1,m_2+s_2} 
	   		  & = & \chi_{i 0 0} \ + \ \Frac{1}{2} \BB \cdot \left( \dd_{i}  \times \TT_{m_1 m_2}  \right) \ + \ \Frac{1}{2} \BB \cdot \left( \dd_{i} \times \TT_{s_1 s_2}  \right) \ + \ f(m_1+s_1,m_2+s_2) \nonumber 
	\\[3mm] & = & \chi_{im_1m_2} \ - \ f(m_1,m_2)  \ + \ \Frac{1}{2} \BB \cdot \left( \dd_{i} \times \TT_{s_1 s_2}  \right) \ + \ f(m_1+s_1,m_2+s_2) \label{eq:first}
	\\[3mm] & = & \chi_{im_1m_2} \ + \ \Frac{1}{2} \BB \cdot \left( \dd_{i} \times \TT_{s_1 s_2} \right) 
	 \ + \ \Frac{1}{2} \BB \cdot \left( \TT_{m_1m_2} \times \TT_{s_1 s_2} \right) \ + \ f(s_1,s_2) \ , \label{eq:last}
\end{eqnarray}
where we made use of \eref{eq:chi_b} for the first two lines and \eref{eq:chi_a} for the last line. 
By comparing \eref{eq:first} and \eref{eq:last}, for consistency we have
\begin{equation} \label{eq:f1}
	f(m_1+s_1,m_2+s_2) \ = \ f(m_1,m_2) \ + \ f(s_1,s_2) \ + \ \Frac{1}{2} \BB \cdot \left( \TT_{m_1m_2} \times \TT_{s_1 s_2} \right) \ .
\end{equation}
After inverting $(m_1,m_2)\leftrightarrow(s_1,s_2)$, the sign of the cross product changes, while the other terms of the previous equation are just swapped
\begin{equation} \label{eq:f2}
	f(m_1+s_1,m_2+s_2) \ = \ f(s_1,s_2) \ + \ f(m_1,m_2) \ + \ \Frac{1}{2} \BB \cdot \left( \TT_{s_1s_2} \times \TT_{m_1 m_2} \right) 
	                   \ = \ f(m_1,m_2) \ + \ f(s_1,s_2) \ - \ \Frac{1}{2} \BB \cdot \left( \TT_{m_1m_2} \times \TT_{s_1 s_2} \right) \ .
\end{equation}
Since, to have a well-defined function $f(m_1,m_2)$, \eref{eq:f1} and \eref{eq:f2} can only differ by a multiple of $\Phi_0$, it follows that
\begin{equation} \label{eq:2d_condition}
	 \BB \cdot \left( \TT_{m_1m_2} \times \TT_{s_1 s_2} \right) \ = \ q \Phi_0  
	\ \ \ \forall m_1, \ m_2, \ s_1, \ s_2 
	     \ \ \ \rightarrow \ \ \ \Phi \ \equiv \ \BB\cdot \left( \TT_1 \times \TT_2 \right) \ = \ q \Phi_0  \ \ \ {\rm with} \ \ \ q\in\mathbb{Z}  \ .
\end{equation}
Therefore, the magnetic flux through the unit cell $\Phi$ must be an integer multiple of the flux quantum $\Phi_0$. 
If $\BB$ lies on the same plane as $\TT_1$ and $\TT_2$, then \eref{eq:2d_condition} is automatically satisfied with $q=0$. 
Otherwise, \eref{eq:2d_condition} determines the minimum admissible magnetic field and its multiples:
\begin{equation} \label{eq:b_condition}
       \BB \ = \ q \ \Frac{\Phi_0}{\bb\cdot \left( \TT_1 \times \TT_2 \right)}\ \bb \  \ \ \ {\rm with} \ q \in \mathbb{Z} \ \ ,
\end{equation}
where $\bb=\BB/|\BB|$ is the direction of the magnetic field. 
Of course, the minimum allowed magnetic field can be decreased by combining more unit cells into a larger supercell.

A simple choice for the function $f(m_1,m_2)$ to satisfy \eref{eq:f1} \textit{modulo} $\Phi_0$ is
\begin{equation}
   f(m_1,m_2) \ = \ \pm\Frac{1}{2} \ \BB \cdot\left( \TT_{m_1}\times\TT_{m_2} \right) \ = \ \pm\Frac{m_1m_2}{2} \ q \ \Phi_0 \ ,
\end{equation}
where the sign is arbitrary, $\TT_{m_1}$ stands for $m_1\TT_1$ and $\TT_{m_2}$ stands for $m_2\TT_2$. 
As a consequence, the gauge function on the lattice is
\begin{equation} \label{eq:chi_chi}
   \chi_{i m_1 m_2} \ = \ \chi_{i 0 0} \ + \ \Frac{1}{2} \ \BB\cdot\left( \dd_i\times\TT_{m_1m_2} \ \pm \TT_{m_1}\times\TT_{m_2} \right)
	                  \ = \ \Frac{1}{2} \ \BB\cdot\left( \dd_i\times\TT_{m_1m_2} \ \pm \TT_{m_1}\times\TT_{m_2} \right)
										\ = \ \Frac{1}{2} \ \BB\cdot\left( \RR_{im_1m_2}\times\TT_{m_1m_2} \ \pm \TT_{m_1}\times\TT_{m_2} \right)
										\ ,
\end{equation}
where we made the arbitrary choice $\chi_{i 0 0}=0$. 
Therefore, together with \eref{eq:b_condition}, the recipe to obtain the Peierls phase in a 2D periodic system is
\begin{eqnarray}  \label{eq:recipe_2d_general} 
  \varphi_{\!\!\!\begin{array}{l} {\scriptstyle  i,m_1,m_2;}\\[-2mm] {\scriptstyle j,n_1,n_2} \end{array}}
	& = & \Frac{\pi}{\Phi_0} \BB\cdot\left[
	      \RR_{im_1m_2}\times\RR_{jn_1n_2} + \RR_{jn_1n_2} \times \TT_{n_1n_2} - \RR_{im_1m_2} \times \TT_{m_1m_2} \pm \TT_{n_1}\times\TT_{n_2} \mp \TT_{m_1}\times\TT_{m_2}
           \right] 
		\\[3mm] & = & \Frac{\pi}{\Phi_0} \BB\cdot\left[
	      \dd_i\times\dd_j + (\dd_i+\dd_j) \times (\TT_{n_1n_2}-\TT_{m_1m_2}) + \TT_{m_1}\times\TT_{n_2} + \TT_{m_2}\times\TT_{n_1}
			\pm \TT_{n_1}\times\TT_{n_2} \mp \TT_{m_1}\times\TT_{m_2}
           \right] \ . \nonumber
\end{eqnarray}
Note that if the magnetic field $\BB$ is not coplanar with the two translation vectors, then from \eref{eq:b_condition} it follows that
\begin{equation} \label{eq:perierls2d_bis}
 \varphi_{i,m_1,m_2;j,n_1,n_2} \ = \ \pi q \ \left[
	\Frac{\bb 
	\cdot\left[
	      \dd_i\times\dd_j + (\dd_i+\dd_j) \times \left( (n_1-m_1)\TT_1+(n_2-m_2)\TT_2\right)\right]} {\bb\cdot \left( \TT_1 \times \TT_2 \right)}\  
				+ \ (m_1 \ \pm \ n_1 ) \ (n_2 \ \mp \ m_2) 
				\right]
     \ .
\end{equation}
Under lattice translations of both $\RR_{im_1m_2}$ and $\RR_{jn_1n_2}$, the first term is manifestly invariant, while the last term always differs by an even integer, which, once multiplied by $\pi q$, gives an irrelevant 2$\pi$-multiple phase.
If, on the contrary, the magnetic field $\BB$ is coplanar with the two translation vectors, then, as mentioned above, its strength does not undergo the condition in \eref{eq:b_condition} and we have
\begin{equation} \label{eq:coplanar_B} 
 \varphi_{i,m_1,m_2;j,n_1,n_2} \ = \ \Frac{\pi}{\Phi_0} \BB\cdot\left[ \dd_i\times\dd_j + (\dd_i+\dd_j) \times (\TT_{n_1n_2}-\TT_{m_1m_2}) \right] \ ,
\end{equation}
which is clearly invariant under lattice translations of both $\RR_{im_1m_2}$ and $\RR_{jn_1n_2}$. 
As detailed in \aref{app:2to1}, apart from an irrelevant redefinition of the gauge function at the edges of the non-periodic region, \eref{eq:coplanar_B} reduces to \eref{eq:peierls_phase_1d} in the case of 1D-systems.

\section{Examples} \label{sec:examples}
In this section, we show some examples of applications of the obtained formulas. 
These examples are not meant to be original nor are they based on very sophisticated models, but just to serve as demonstration of the application of the discussed theoretical formulation. 
We first consider the case of a metallic carbon nanotube (CNT) in a homogeneous magnetic field. 
Such a configuration has been widely investigated in the literature~\cite{Ajiki1993,Saito1994,Ajiki1996,Roche2000,Nemec2006,Charlier2007}, both in the tight-binding and the $k\cdot p$ descriptions. 
Depending on the angle between the magnetic field and the CNT axis, different phenomena occur, such as the formation of Landau levels (LLs) or the opening of a gap due to the Aharonov-Bohm effect~\cite{Ajiki1993}. 
The second example is a six-terminal Hall bar obtained in a two-dimensional electron gas (2DEG). Here, the presence of chiral edge states and of localized states around impurities is central for the observation of extended quantized Hall resistance plateaus. 
Finally, we consider a 2D graphene superlattice of Gaussian bumps. This system has attracted much interest thanks to the possibility of inducing pseudomagnetic fields by controlling strain~\cite{Guinea2009,Levy2010,Banerjee2020}.

\subsection{Metallic carbon nanotubes in a magnetic field}\label{subsec:cnts}
\begin{figure}[!tb] 
	\centering
		\resizebox{9.5cm}{!}{\includegraphics{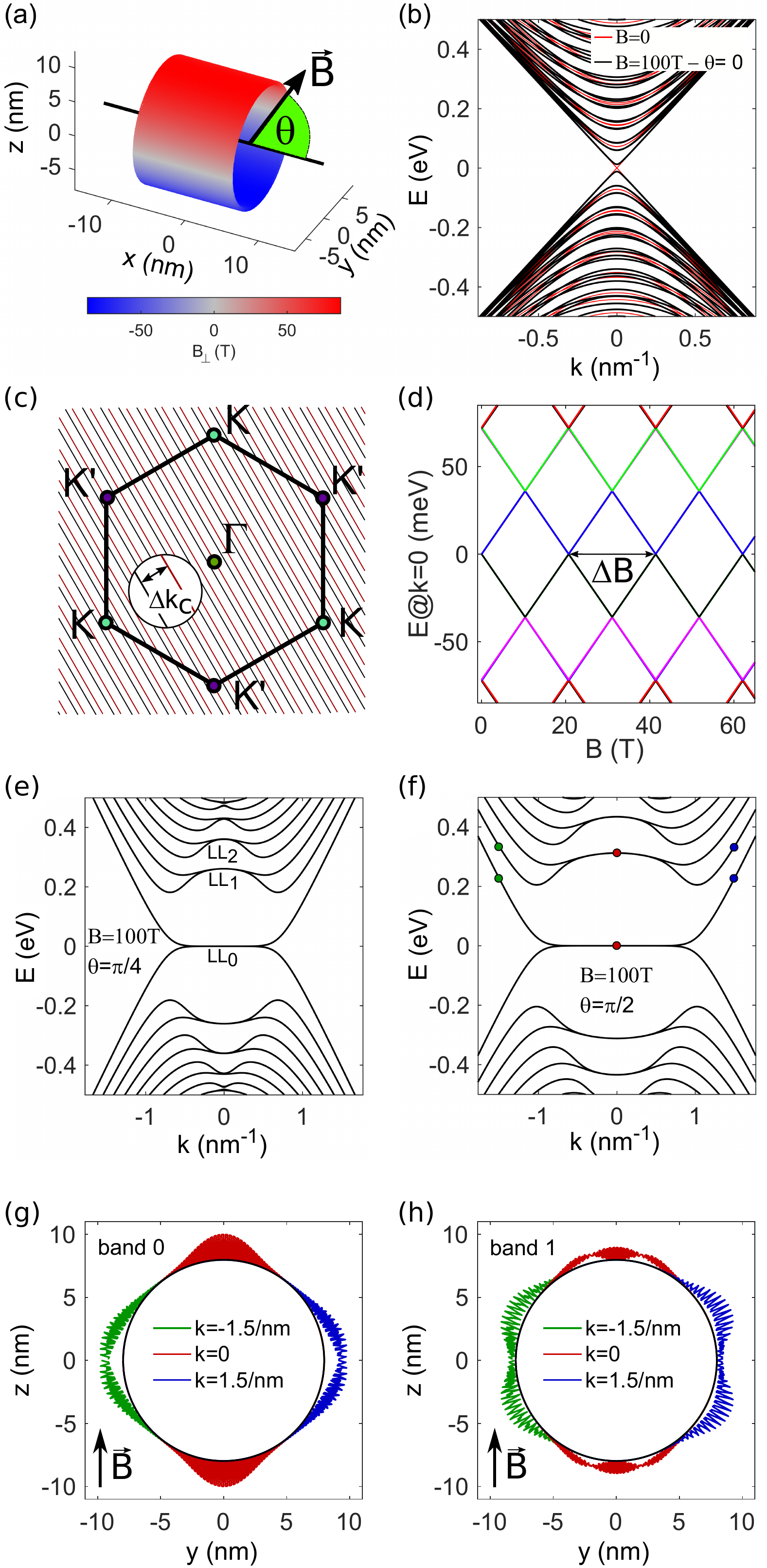}}
		\caption{\label{fig_cnt_bands}
		(a) CNT with chirality (204,0), corresponding to a radius of about 8~nm and a circumference of about 50~nm. The magnetic field forms an angle $\theta$ with the nanotube axis. 
The color scale indicates the normal component of the magnetic field $B_\perp$, which is maximum and with opposite sign on the top and the bottom regions of the CNT, while it vanishes at the sides of the CNT. 
(b) Band structure in the absence (red lines) and in the presence (black lines) of a 100~T magnetic field along the CNT axis ($\theta=0$). 
		(c) Slicing of the hexagonal Brillouin zone of graphene with quantized transverse wave vector. The Dirac cones are at the K and K' points at the corner of the Brillouin zone. The cases $B=0$ and $B\neq 0$ correspond to the red and black lines, respectively. 
		(d) Energy of the first low-energy bands at $k=0$ as a function of the magnetic field along the CNT axis. 
		(e) Band structure for $B=100$~T and $\theta=\pi/4$. The first LLs with $E\geq0$ are indicated. 
		(f) Band structure for $B=100$~T and $\theta=\pi/2$. (g) Probability density for the three states indicated by green, blue, and red dots at $k=-1.5$/nm, $k=0$, and $k=1.5$/nm, respectively, along the lowest band in (f). 
		(h) Same as (g) for states along the first band.
		}
\end{figure}

We consider a CNT with chiral indices $(n=204,m=0)$, which corresponds to a metallic zigzag CNT with a radius $R\approx8$~nm and a circumference of about 50~nm~\cite{Charlier2007}. 
We describe the CNT by a tight-binding Hamiltonian with a single $p_z$ orbital {\it per} carbon atom and first-nearest-neighbor coupling 
\begin{equation} \label{eq:hamiltonian_cnt}
	H \ = \ t_0 \ \sum_{\langle ij\rangle} \ket{\phi_i}\bra{\phi_j} \ ,
\end{equation}
where $t_0=-2.7$~eV is the hopping parameter, $\ket{\phi_i}$ corresponds to the $p_z$ orbital of the carbon atom with index $i$, and $\langle i j\rangle$ indicates that the atoms with indices $i$ and $j$ are first neighbors. 
We make use of \eref{eq:peierls_phase_1d} with unit vector $\tt$ along $\hat{\vec{x}}$ to take into account the presence of a homogeneous magnetic field $\BB$ that forms an angle $\theta$ with the CNT axis, see \fref{fig_cnt_bands}(a). 
The color scale in this figure represents the projection of the magnetic field $B_\perp$ along the normal to the CNT curved surface. 
$B_\perp$ is maximum at the top of the CNT, it decreases to 0 at its sides, and changes sign to reach its maximum negative value on the bottom of the CNT. 
This quantity corresponds to the varying orthogonal magnetic field locally experienced by electrons, and it is useful for the interpretation of the results, as we will see later on.    

In the absence of magnetic field, see the red lines in \fref{fig_cnt_bands}(b), the band structure presents a linear dispersion at the charge neutrality point, which corresponds to the graphene Dirac point, and quantized subbands whose spacing is inversely proportional to the CNT diameter. 
The K and K' points of the hexagonal Brillouin zone of 2D graphene are folded at $k=0$ and the bands are valley degenerate. 

When a $B=100$~T magnetic field is along the ribbon axis ($\theta=0$), a small gap opens and the degeneracy lift of the other bands occurs, see the black lines in \fref{fig_cnt_bands}(c), as observed in \cref{Ajiki1993}. 
Remarkably, these changes occur despite the fact that $B_\perp=0$, i.e., the magnetic flux vanishes through the carbon hexagons of the CNT lattice. 
To understand this behavior, we consider the so-called \emph{zone-folding approximation}~\cite{Hamada1992,Saito1992}.
The low-energy band structure of a CNT can be obtained from the energy dispersion of 2D graphene and considering that the transverse component of the wave vector $k_c$, i.e., orthogonal to the CNT axis, must be quantized to ensure the single valuedness of the electron wave function around the CNT circumference. 
Indeed, according to the Bohr-Sommerfeld quantization condition, we have
\begin{equation} \label{eq:sommerfeld}
  \oint \ppi\cdot d\vec{r} \ = \ \oint \left( \hbar\vec{k} \ + \Frac{e}{c}\vec{A}\right)\cdot d\vec{r} \ = \ \hbar k_c 2\pi R \ + \ \Frac{e}{c}\Phi(\vec{B}) \ = \ 2\pi q \hbar
	\ \ \ \rightarrow \ \ \
	k_c \ = \ \Frac{1}{R} \ \left( q \ - \ \Frac{\Phi(\vec{B})}{\Phi_0}  \right) \ ,
\end{equation}
where the integral is performed along the CNT circumference, $\Phi(B)=\pi R^2 B$ is the magnetic flux through the CNT section and $q$ is an integer number.
The bands are thus obtained by slicing the hexagonal Brillouin zone of 2D graphene along lines that are parallel to a line with angle $\pi/3$ and passing through the $\Gamma$ point, and spaced by $q/R$. 
For the considered CNT chirality ($n$ is a multiple of 3 and $m=0$) and for $B=0$, one of these lines passes exactly at the K and K’ points, see the red lines in \fref{fig_cnt_bands}(c), where the graphene Dirac cones are located. 
Therefore, the bands are valley degenerate with the K and K' points folded at $k=0$, and the system is metallic, with linear dispersion at the charge neutrality point.  
When a magnetic field is present, the shift in $k_c$ by $\Delta k_c=\Phi(\vec{B})/(R\Phi_0)$ results in a shift of the slicing of the Brillouin zone, see the black lines in \fref{fig_cnt_bands}(c). 
The slices do not pass anymore through the K and K' points and they are not equidistant from them, and then we observe the opening of the band gap and the degeneracy removal for the other bands. 
\Fref{fig_cnt_bands}(d) shows the energy of the first low-energy (around $E=0$) bands at $k=0$ as a function of the magnetic field $B$ along the CNT axis. We can clearly observe the linear shift of the energies and a periodicity $\Delta B = \Phi_0 / (\pi R^2) \approx 20.7$~T, which corresponds exactly to integer multiples of the quantum magnetic flux through the CNT section, as expected from \eref{eq:sommerfeld}.

When the angle between the $B=100$~T magnetic field and the CNT axis is increased to $\theta=\pi/4$ and then to $\theta=\pi/2$, see \ffref{fig_cnt_bands}(e) and \ref{fig_cnt_bands}(f), the usual series of LLs develops~\cite{Nemec2006} since the magnetic length $\ell=\sqrt{\hbar c/(eB)}\approx 2.5$~nm is shorter than the CNT circumference. 
However, the LLs are dispersive and not flat as in graphene. 
This is due to the varying orthogonal component of the magnetic field $B_\perp$, which is maximum on the top and bottom regions and vanishes on the CNT sides, where the magnetic field is tangential, see \fref{fig_cnt_bands}(a). 
We also remark that LLs are degenerate due to the invariance of the system composed by CNT plus magnetic field under the symmetry operation given by the combination of a 
$C_2$ rotation around the CNT axis with a mirror reflection on a plane passing through the CNT axis and the CNT sides, where $B_\perp$ vanishes. 
Such symmetry corresponds to having two equivalent halves of CNT with opposite components of the orthogonal magnetic field. 

We further investigate the electronic structure by looking at the spatial probability distribution of the states corresponding to the colored dots at $k=0$ and $k=\pm1.5$/nm on the first two bands in \fref{fig_cnt_bands}(f), reported in \fref{fig_cnt_bands}(g) and \fref{fig_cnt_bands}(h), respectively. 
At $k=0$ (red dots and lines), we observe the Landau states corresponding to LL0 and LL1 at the top and the bottom of the CNT. 
As expected, when changing $k$ toward positive or negative values, the Landau state centers move toward the right or left side of the CNT, and, according to the dispersive energy dispersion, the LLs acquire a group velocity whose direction depends on the sign of the nonvanishing gradient of $B_\perp$. 
When the states reach the region of vanishing or small $B_\perp$, see the green and blue lines in \ffref{fig_cnt_bands}(g) and \ref{fig_cnt_bands}(h), the LLs cannot form since the resulting magnetic length is longer than the CNT circumference. 
As a consequence their energy dispersion becomes almost linear. 
Analogously to what was observed for edge states in Hall bars discussed below in \sref{sec:IQHE}, these states entail a spatial chirality, in the sense that they show opposite group velocities at the opposite sides of the Brillouin zone, see the slope of the bands in correspondence of green and red dots in \ffref{fig_cnt_bands}(e) and \ref{fig_cnt_bands}(f), and are located at opposite sides of the CNT, see the green and red lines in \ffref{fig_cnt_bands}(g) and \ref{fig_cnt_bands}(h).    
This phenomenon can be conveniently described in term of semiclassical \emph{snake states}, which originate from a sign inversion of the magnetic field and whose group velocity direction depends on the magnetic field gradient, as detailed in \cref{Mueller1992}, and are also observed in scrolled graphene~\cite{Cresti2012}.

\subsection{Integer quantum Hall effect in a 2DEG Hall bar}\label{sec:IQHE}
\begin{figure}[!tb] 
	\centering
		\resizebox{9cm}{!}{\includegraphics{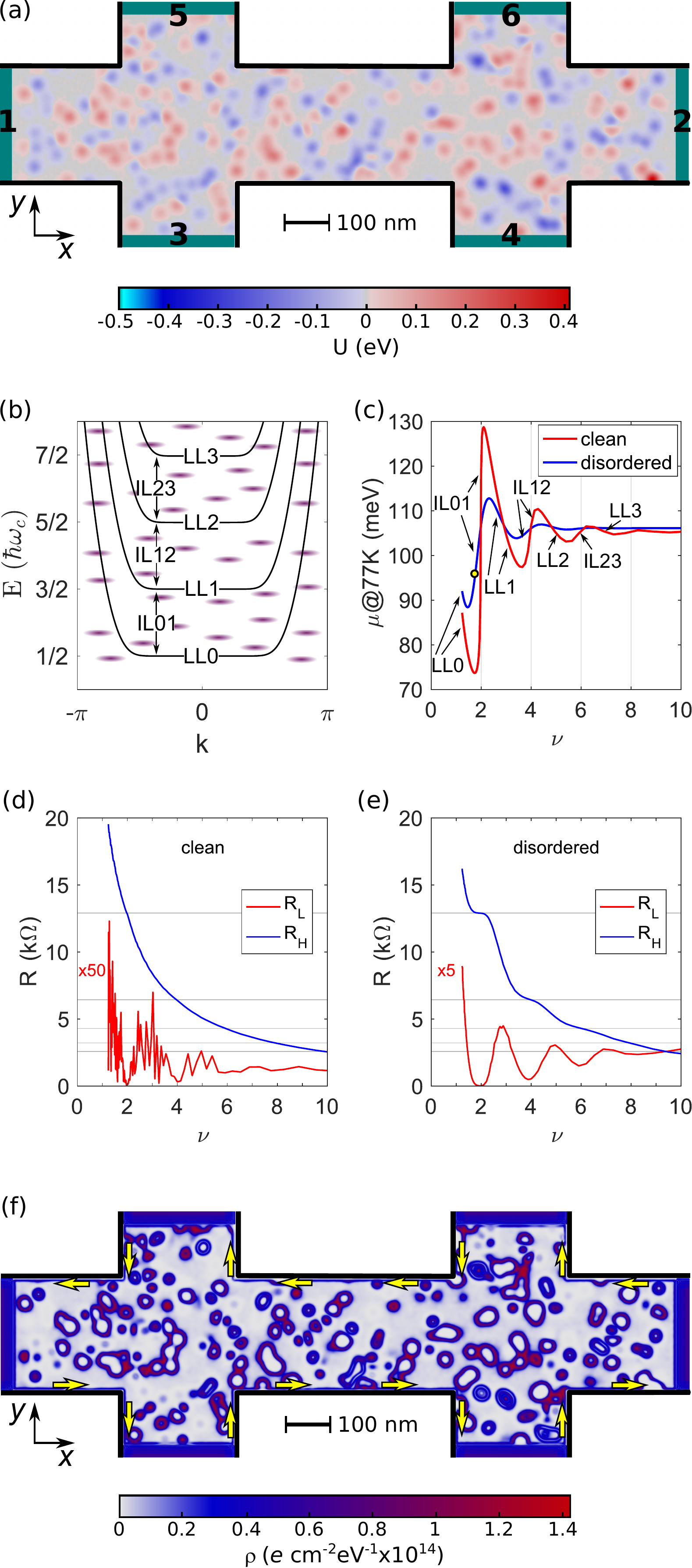}}
		\caption{\label{fig_hall_bar}(a) Geometry of the six-terminal Hall bar and background potential energy. The terminal numbers from 1 to 6 are indicated. (b) Typical band structure of a 2DEG ribbon in high magnetic field. It is characterized by the presence of LLs and edge states in the interlevel regions. The presence of localized states induced by disorder is schematically indicated by purple energy levels. (c) Chemical potential as a function of the filling factor for the Hall bar in (a) in the cases of clean and disordered samples, for charge density $\sigma=3\times10^{13}$~$e$/cm$^2$ and temperature $T=77.36$~K. (d) Longitudinal and Hall resistances for the clean sample. The horizontal grey lines correspond to the resistance $h/(2e^2)$ and its submultiples. (e) Same as (d) for the disordered sample. (f) Surface density of electrons {\it per} unit of energy in the disordered sample for $B=70$~T, $T=77.36$~K, and $\mu=96.15$~meV, corresponding to the dot at $\nu=1.77$ in (c).}
\end{figure}

We consider here a Hall bar etched in a 2DEG obtained at the heterojunction between GaAs and AlGaAs. 
The geometry, reported in \fref{fig_hall_bar}(a), consists of six terminals with width 250~nm and a central bar with same width and a length of 1.5~$\mu$m, where disorder may be present. 
The current probes correspond to the terminals with indices 1 and 2, while the voltage probes to the terminals with indices 3, 4, 5, and 6. 
To mimic metallic contacts, all the terminals are electrostatically doped by superimposing a constant negative potential energy $-0.5$~eV. 
Note that the size of the Hall bar is much smaller than that of typical experimental bars for metrological application~\cite{Jeckelmann2001}. 
This choice, whose motivation is to limit the computational burden, does not affect the physical interpretation of the results, which are in any case intended as demonstrative in this article.  
We consider two cases: a clean Hall bar with very weak disorder, and a disordered Hall bar where long-range energy potential centers are included to mimic charged impurities. 
The contrast between the two cases is interesting, because the width of the quantized Hall resistance plateaus is expected to be larger in disordered than in ultraclean samples. 
To model disorder, we consider $N$ randomly distributed impurities with Gaussian profile
\begin{equation}
	V(\vec{r}) \ = \ \sum_{j=1}^N v_j \ \exp\left(-\Frac{\left| \vec{r}-\vec{r}_j \right|^2}{2\lambda^2}  \right) \ ,
\end{equation}
where $\{v_j\}$ are randomly chosen in the range $[-s,s]$, with $s$ the strength of the disorder, $\{\vec{r}_i\}$ are the random positions of the potential centers, and $\lambda$ is the potential range.

The weak disorder is obtained by first generating a random potential with parameters $\lambda=2$~nm and a very high density of centers, $10^{14}$~cm$^{-2}$, and then by renormalizing it to have a root mean square value of 1~meV.
The long-range disorder, superimposed to the weak disorder considered above, is generated by setting $\lambda=10$~nm, $s$=150~meV, and impurity density $7.5\times 10^{10}$~cm$^{-2}$. 
A realization of the resulting potential is illustrated in \fref{fig_hall_bar}(a).

The tight-binding-like Hamiltonian is obtained by discretizing the continuous effective mass Hamiltonian over a square lattice with parameter $a$~\cite{Datta1995}:
\begin{equation}
	H \ = \ \Frac{|\vec{p}|^2}{2m^*} 
	\ \ \ \ \rightarrow \sum_i \ket{\phi_i}(4t+U_i)\bra{\phi_i} \ - \ t \sum_{\langle ij\rangle} \ket{\phi_i}\bra{\phi_j} \ ,
\end{equation}
where $m^*=0.068~m_e$ is the electron effective mass, $t=\hbar^2/(2m^*a^2)$ is the hopping parameter, $\{\ket{\phi_i}\}$ is an (incomplete) basis of localized states at the sites $\{i\}$ of the lattice, $U_i$ is the superimposed energy potential (including disorder and contact electrostatic doping) at site $i$, and $\langle ij\rangle$ indicates that the sum is limited to first-nearest-neighbor sites. 
We choose $a=1$~nm, which is a good compromise between accuracy (because it is smaller than the long-range disorder range and the typical magnetic length here considered) and computational burden. 

The presence of a homogeneous orthogonal magnetic field is considered according to the recipe of \eref{eq:peierls_phase_1d} and by taking into account the different orientations of the semi-infinite terminals, i.e., a unit vector $\tt$ oriented along the $x$~axis for contacts 1 and 2 and along the $y$~axis for contacts 3, 4, 5, and 6. 
We briefly recall the band structure of a periodic 2DEG ribbon in a high magnetic field, which is shown in \fref{fig_hall_bar}(b). 
First of all, we observe flat LLs with energy \mbox{$E_n(B)=(n+1/2)\hbar\omega_c$}, where $\omega_c=eB/(m^*c)$ is the cyclotron frequency, and each with degeneracy $\Phi(B)/\Phi_0$, where $\Phi(B)$ is the magnetic flux through the whole 2DEG. 
At energies between LLs, which we call inter-level (IL) regions, dispersive bands are present, which correspond to edge states. 
Such edge states are chiral, in the sense that electrons flow along them in opposite directions at opposite edges~\cite{Cresti2004,Cresti2006}, as follows from the opposite slope of the dispersive bands at opposite sides of the Brillouin zone. 
In contrast to the bulk of the wire, which is an insulator since there are no states in the IL energy regions, these states are perfectly conducting as long as disorder is not strong enough to couple states at opposite edges. 
This phenomenon is at the origin of the Hall resistance quantization~\cite{Buttiker1988}. 
As mentioned above and detailed below, disorder can induce localized states around the impurities, which are illustratively represented by the purple energy levels in \fref{fig_hall_bar}(b).

The Green's function approach and the Landauer-B\"{u}ttiker formalism allow us to calculate, starting from the above Hamiltonian, the transmission coefficients between the different terminals as a function of the electron energy in the linear response regime~\cite{Sumetskii1991} and the surface electron density as a function of the chemical potential $\mu$ and the temperature $T$~\cite{Areshkin2010}. 
In our simulations~\cite{Cresti2003}, the temperature just entails a smearing of the Fermi-Dirac distribution function, while no electron-phonon coupling is taken into account. 
We set the average surface charge density in the Hall bar to $\sigma=3\times10^{13}$~$e$/cm$^2$ and determine the chemical potential $\mu$ at temperature $T=77.36$~K, corresponding to that of liquid nitrogen, and for different magnetic fields up to 100~T. 
We choose such a high electron surface density to be able to explore the region $\nu>2$ at relatively high magnetic fields, for which the magnetic length is comparable to or smaller than the system size.  
The result is reported in \fref{fig_hall_bar}(c) in terms of the filling factor $\nu=\Phi_0 \sigma/B$. 
In the case of very weak disorder (red line), we observe the typical oscillations of the chemical potential, which jumps on different LLs when $\nu$ is a multiple of 2 (we consider spin degeneracy) by rapidly filling the edge states (regions indicated by IL01, IL12, and IL13 in the figure).  
For different values of $\nu$, the chemical potential is roughly pinned to the LLs and increases almost linearly with the magnetic field. These regions are indicated by LL0, LL1 and LL2 for the first three LLs. 
The $\nu$~extension of IL regions compared to that of the LL regions depends on the ratio of the density of states corresponding to the edge states over that corresponding to the bulk LLs. 
At larger $\nu$, the width of the semiclassical cyclotron orbit, and then of the edge states, is larger, thus entailing an increased extension of the IL regions. 
This is particularly visible in our small-size system. 
In the case of stronger Gaussian disorder (blue line), the density of states is expected to broaden due to the formation of localized states. The result is that the transition between LLs is much slower and the chemical potential stays in the IL regions, where also chiral edge states are present, for a much wider interval of magnetic field. 
We thus expect to observe wider Hall resistance plateaus in disordered samples compared to clean or very weakly disordered samples, as also observed experimentally. 

To verify this prediction, we plot in \ffref{fig_hall_bar}(d) and \ref{fig_hall_bar}(e) the longitudinal and Hall resistances as a function of $\nu$. 
To obtain them from the transport coefficients, we first determine the terminal potentials $\{V_n\}$ by imposing a source drain current of $I_{\rm sd}=0.1~\mu$A and a vanishing current on the voltage probes. 
Then, we calculate the longitudinal resistance as $R_{\rm L}=(V_6-V_5)/I_{\rm sd}$ and the Hall resistance as $R_{\rm L}=(V_6-V_4)/I_{\rm sd}$. 
In the case of clean graphene, \fref{fig_hall_bar}(d) shows that the Hall resistance is not quantized and that the longitudinal resistance is rather small and only barely presents some dips in correspondence of integer even $\nu$. 
The quantum Hall effect is thus not observable in this case and at the considered temperature of 77.36~K. 
In contrast, for the disordered sample, \fref{fig_hall_bar}(e) clearly shows the Hall resistance plateau with value $h/(2e^2)\approx 12.9$~k$\Omega$ for $\nu=2$, where the longitudinal resistance vanishes. For $\nu=4,~6...$, only flexes of the Hall resistance are observable at the considered temperature. 
However, the longitudinal resistance clearly shows the corresponding minima.  

Finally, to provide a visual picture of the mechanisms at the origin of resistance quantization and the pinning of the chemical potential at energies between the LLs, \fref{fig_hall_bar}(f) reports the surface density of electrons {\it per} unit of energy $\rho$ at $B=70$~T, $T=77.36$~K, and chemical potential $\mu=96.15$~meV, see the dot at $\nu=1.77$ in \fref{fig_hall_bar}(c). 
This quantity is obtained by deriving the total electron density~\cite{Areshkin2010} with respect to the chemical potential.
We can clearly observe (i) a large density of electrons on the doped region of the terminals, (ii) chiral edge states at the edges of the bar, with indication of the anticlockwise direction of their group velocity, which is opposite at opposite edges (see the yellow arrows), and (iii) the localized states in the insulating bulk, which turn around the Gaussian impurities in clockwise or anticlockwise direction (depending on the impurity potential sign) and largely contribute to the density of states and then to the Fermi level pinning, but not to transport. 
This result completely agrees with the edge state model proposed in \cref{Buttiker1988}. 
Note that, in the simple model here adopted, we did not consider edge reconstruction or electron-electron or electron-phonon coupling, which are out of the scope of the present paper.

\subsection{Periodic bumped graphene superlattice}
\begin{figure}[!tb] 
	\centering
		\resizebox{14cm}{!}{\includegraphics{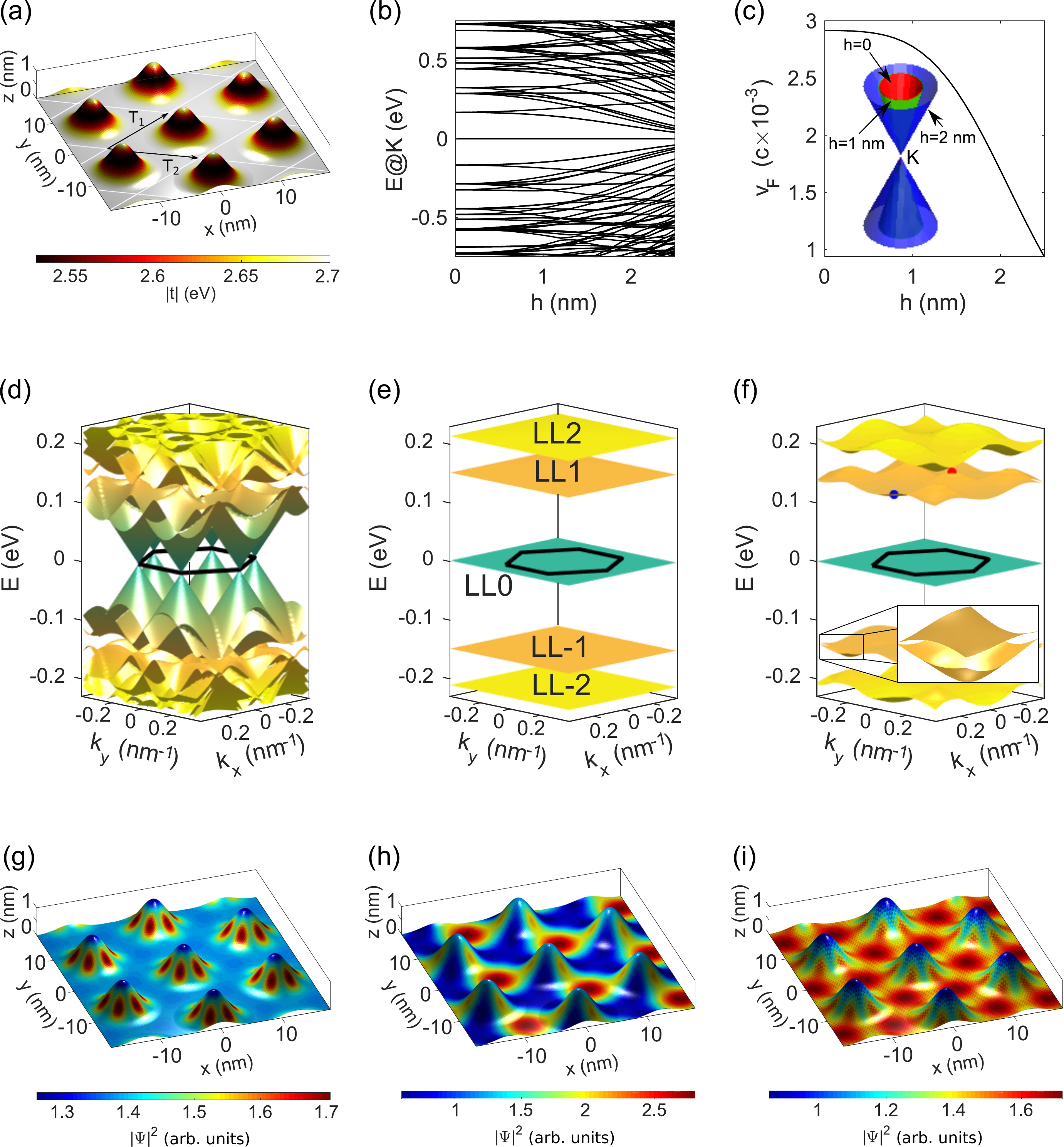}}
		\caption{\label{fig_2dgraphene}
		(a) Graphene profile in the presence of the Gaussian deformation of \eref{eq:gaussian_profile} with $h=1$~nm and $b=2.92$~nm. The unit cell is periodically repeated with translation vectors $\TT_1$ and $\TT_2$. The color scale indicates the local average value of the first nearest neighbor hopping parameters for each atom according to \eref{eq:TB_strain}. 
		(b) Energy levels at the point K of the Brillouin zone as a function of the parameter $h$. 
		(c) Fermi velocity as a function of the parameter $h$. The inset shows the Dirac cones at the point K for $h=0$, 1 and 2~nm. 
		(d) Energy bands close to $E=0$ for $h=1$~nm and $B=0$. The first Brillouin zone is indicated by a black hexagon. 
		(e) Same as (d) for $h=0$ (no deformation) and $B=22.74$~T. The LLs are indicated. 
		(f) Same as (d) for $B=22.74$~T. The inset shows a zoom of the indicated area. 
		(g) Probability density for an eigenvector $\Psi$ of the Hamiltonian close to the $K$ point of the Brillouin zone, for $h=1$~nm and $B=0$. 
		(h) Same as (g) for $h=1$~nm and $B=22.74$~T at the M point of the Brillouin zone, corresponding to the red dot in (f). 
		(i) Same as (h) for a wave vector in the middle between the K and the $\Gamma$ points, corresponding to the blue dot in (f).
		}
\end{figure}

Bumps are often observed in graphene samples and can be artificially induced, for example, by pinning graphene on patterned substrates~\cite{Kusminskiy2011,Georgiou2011} or by thermally inducing a buckling transition~\cite{Mao2020}. 
In the absence of magnetic field, such a system has been widely investigated in the literature~\cite{Moldovan2013,Schneider2015,CarrilloBastos2014,Settnes2016,Tran2020}, with focus on both electronic structure and transport properties in 2D graphene and graphene ribbons or dots.
Here, we focus on the case of periodic bumps with a Gaussian profile similar to that of \cref{Moldovan2013}
\begin{equation} \label{eq:gaussian_profile}
	 z(r) \ = \ \left\{ \begin{array}{ll} h \ \Frac{e^{-r^2/b^2}-e^{-R^2/b^2}}{1-e^{-R^2/b^2}} & \ \ \ {\rm for} \ \ \ r\leq R \\[5mm]
                                        0  & \ \ \ {\rm for} \ \ \ r > R
				       \end{array}\right.
\end{equation}
where $r$ is the distance from the deformation center, $h$ is the maximum height of the deformation, $b$ is the width parameter and $R=3b/\sqrt{2}$ is a cutoff radius, which contains the 99.7\% of the Gaussian. 
Following \cref{Moldovan2013}, we choose $h$ within the range $0-2.5$~nm, $b=2.9227$~nm, and then $R=6.2$~nm. 
We consider the smallest triangular unit cell multiple of the graphene unit cell that contains the deformation. 
It has translation vectors $\TT_{1,2}=[12.5493,\pm 7.2453,0]$~nm and contains 6962 carbon atoms. 
The resulting structure is shown in \fref{fig_2dgraphene}(a) for $h=1$~nm. 
With this choice, the Brillouin zone keeps the original hexagonal shape and, in our specific case, the original K and K' points turn out to be folded into the K and K' of the new Brillouin zone. 

We adopt the same tight-binding Hamiltonian as for CNTs with a single $p_z$ orbital {\it per} atom with first-nearest-neighbor coupling: 
\begin{equation} \label{eq:hamiltonian_carbon}
	H \ = \ \sum_{\langle ij\rangle}\ t_{ij} \ \ket{\phi_i}\bra{\phi_j} \ ,
\end{equation}
where, to take into account the strain resulting from the deformation, the hopping parameter $t_{ij}$ depends on the inter-atomic distance $d_{ij}$ between the atoms $i$ and $j$ as~\cite{Pereira2009}  
\begin{equation} \label{eq:TB_strain}
	 t_{ij}\ = \ t(d_{ij}) \ = \ t_0 \ \exp\left[ -\beta \ \left(\Frac{d_{ij}}{a}\ -\ 1 \right) \right] \ ,
\end{equation}
where $t_0=-2.7$~eV, $a=0.1418$~nm is the unstrained inter-atomic distance and $\beta=3$. 
\Fref{fig_2dgraphene}(a) shows the value of the hopping parameter averaged over the three nearest-neighbor couplings of each carbon atom. 
Graphene is almost unstrained on the tops of the bumps and at their bases, while it is maximally tensile strained on the sides of the bumps, where the hopping parameter can be considerably decreased.

We first examine the effect of the deformation at the K point. 
\Fref{fig_2dgraphene}(b) shows the eigenvalues of the Hamiltonian as a function of the parameter $h$. 
Independently of it, an $E=0$ level is always present. 
This means that the Dirac point is always at K and a gap does not open. 
The other energy levels, which result from the folding of the original graphene Brillouin zone into the new smaller one, are mainly observed to decrease when increasing $h$, as a consequence of the average decrease of the hopping elements in the Hamiltonian. 
However, a few levels are observed to increase. 
This behavior might follow the rise of a pseudomagnetic field induced by the strain~\cite{Guinea2009}. 
Indeed, in the low-energy continuous approximation of graphene represented by a Dirac equation for each of the two valleys, the strain effect encoded by \eref{eq:TB_strain} translates into a pseudomagnetic field, which is opposite in the two valleys in accordance with the time-reversal symmetry. 
In the case of a uniform pseudomagnetic field, generated by particular strain profiles, pseudo LLs are predicted~\cite{Guinea2009} and experimentally observed~\cite{Levy2010} to rise. 
In our case, the pseudomagnetic field is inhomogeneous with a trigonal symmetry~\cite{Schneider2015} due to the spatially varying strain. 
However, the increased value of the pseudomagnetic field for large deformations can justify the condensation of eigenvalues in the $E=0$ level and in higher-energy pseudo LLs, as discussed in \cref{Moldovan2013}.  

As shown in \fref{fig_2dgraphene}(c), the increase of $h$ also entails a nonlinear decrease of the Fermi velocity $v_{\rm F}$, which corresponds to an increase of the Dirac cone angle at K and K', see the inset.
The complete low-energy  band structure for $h=1$~nm and $B=0$ is shown in \fref{fig_2dgraphene}(d). 
The Dirac cones at the corners of the Brillouin zone, indicated by a black hexagon, as well as the folding of the original Brillouin zone are clearly visible. 
The presence of weakly dispersive energy band regions is interesting for the possible observation of the effects of electron-electron interactions, as discussed in \cite{Mao2020}.
\Fref{fig_2dgraphene}(g) reports the probability density for an eigenvector of the Hamiltonian with wave vector close to the K point. 
As already observed in the literature~\cite{Schneider2015}, a flower-like structure appears on the side of the bump, which reflects the trigonal symmetry of the geometry and where the petals can be shown to be alternatively polarized on the two graphene sublattices.

We now include an orthogonal magnetic field by making use of the recipe of \eref{eq:recipe_2d_general} or, equivalently, \eref{eq:perierls2d_bis}. 
Given the size of the unit cell, the minimum magnetic field we can consider along the $z$~direction is $B=22.7424$~T. 
Note that, for the sake of simplicity, we consider spin degeneracy and do not include the Zeeman Hamiltonian term. 
In the absence of deformation, i.e., for $h=0$, \fref{fig_2dgraphene}(e) shows that flat LLs appear with the expected energy~\cite{McClure1956}
\begin{equation}\label{eq:LLs}
	E_n \ = \ \pm 3\sqrt{3} \ \Frac{a|t|}{\sqrt{\Phi_0}} \ \sqrt{B} \sqrt{n} \ = \ 
	\pm\sqrt{2e\hbar v_{\rm F}}  \ \sqrt{B} \sqrt{n} \ \approx \ \pm 31.6565 \ \sqrt{B(T)} \ \sqrt{n} \ {\rm meV} \ . 
\end{equation} 
Each level is fourfold degenerate as a consequence of valley and spin degeneracy, and of the fact that the flux of the magnetic field through the unit cell is exactly $\Phi_0$~\cite{Goerbig2011}. 
\Fref{fig_2dgraphene}(f) shows the energy bands in the presence of periodic deformations with $h=1$~nm. 
While the LL0 is still at $E=0$, the other LLs have an average lower energy compared to the case $h=0$ due to the average decrease of the Fermi velocity, see \fref{fig_2dgraphene}(b), which enters \eref{eq:LLs}.
Except for LL0, the other LLs are not flat due to the local variation of the hopping energy (or Fermi velocity) over a spatial range comparable to the magnetic length $\ell\approx 5$~nm. 
As a consequence, the energy will be higher when the eigenstates are mainly located in the unstrained areas, and lower when the eigenstates are also located on the sides of the Gaussian bumps. 
To corroborate this explanation, \ffref{fig_2dgraphene}(h) and \ref{fig_2dgraphene}(i) show the density probability for two eigenstates corresponding to a minimum and a maximum of the third band, indicated by the red and blue dots, respectively, in \fref{fig_2dgraphene}(f). 
\Fref{fig_2dgraphene}(h) confirms that the state with higher energy is mainly located in the regions within the bumps, where the strain is small. Conversely, \fref{fig_2dgraphene}(i) shows that the state with lower energy is also distributed on the sides of the bumps, where the strain is larger.

Finally, the inset of \fref{fig_2dgraphene}(f) reveals that the valley degeneracy is lifted. 
This is consequence of the combined action of magnetic and pseudomagnetic fields~\cite{Kim2011}. 
Indeed, while the magnetic field is the same for the two valleys, the pseudomagnetic field is opposite. 
Their vector sum is thus different in the two valleys, which corresponds to a different effective magnetic field and then to different energy levels.

\section{Conclusions} \label{sec:conclusions}
We provided explicit and practical recipes for the Peierls phase factors to introduce in tight-binding-like Hamiltonians in order to account for a homogeneous magnetic field. 
A proper choice of the gauge allowed us to obtain \eref{eq:peierls_phase_1d} for {\it quasi}-1D periodic systems, including Hall bars with differently oriented terminals, and \eref{eq:recipe_2d_general} and \eref{eq:b_condition} for 2D periodic systems. 

We illustrated some simple but relevant examples of application of the formulas. 
In particular, we investigated the impact of a high magnetic field on the electronic structure of a metallic CNT. 
Different angles between the magnetic field and the nanotube axis revealed a rich physics ranging from Landau states to the Aharonov-Bohm effect.
Then, we simulated the integer quantum Hall effect in 2DEG bars. We highlighted the role of disorder in allowing the observation of resistance plateaus. 
Finally, we considered the case of periodic 2D graphene with Gaussian bumps, where the strain was found to make LLs dispersive and to lift the valley degeneracy.

Our results represent a practical tool for the simulation of electronic and transport properties of mesoscopic systems in the presence of magnetic fields.


\appendix

\section{Mathematical details of the Peierls phase derivation} \label{app:peierls}
In this appendix, following \cref{Luttinger1951}, we provide the mathematical details of the approximation $\nabla G_n(\rr)\approx \AA(\rr)$ used in \sref{sec:peierls} for the derivation of \eref{eq:peierls_phase}. 
We start by introducing the vectorial calculus identity
\begin{equation}
		\nabla(\vec{a}\cdot\vec{b}) \ = \ (\vec{a}\cdot\nabla)\vec{b} \ + \ (\vec{b}\cdot\nabla)\vec{a} \ + \  \vec{a}\times(\nabla\times\vec{b}) \ + \  \vec{b}\times(\nabla\times\vec{a}) \ ,
\end{equation}
which can be verified by direct inspection
\begin{eqnarray}
		\partial_{x_i} a_j b_j & = &       a_j \partial_{x_j} b_i
		                             \ + \ b_j \partial_{x_j} a_i
																 \ + \ \epsilon_{ijk} \epsilon_{kmn} a_j \partial_{x_m} b_n
		                             \ + \ \epsilon_{ijk} \epsilon_{kmn} b_j \partial_{x_m} a_n 
									 \\[3mm] & = &	     a_j \partial_{x_j} b_i
		                             \ + \ b_j \partial_{x_j} a_i
																 \ + \ (\delta_{mi}\delta_{nj}-\delta_{mj}\delta_{ni}) a_j \partial_{x_m} b_n
		                             \ + \ (\delta_{mi}\delta_{nj}-\delta_{mj}\delta_{ni}) b_j \partial_{x_m} a_n 			 									 
									 \\[3mm] & = &	     \cancel{a_j \partial_{x_j} b_i}
		                             \ + \ \bcancel{b_j \partial_{x_j} a_i}
																 \ + \ a_j \partial_{x_i} b_j \ - \ \cancel{ a_j \partial_{x_j} b_i}
		                             \ + \ b_j \partial_{x_i} a_j \ - \ \bcancel{b_j \partial_{x_j} a_i} 			 
									 						\ ,
\end{eqnarray}
where we assume the sum over repeated indexes, exploit the identity $[\vec{a}\times\vec{b}]_i=\epsilon_{ijk} a_j b_k$, with $\epsilon_{ijk}$ the Levi-Civita antisymmetric symbol, and consider that $\epsilon_{ijk} \epsilon_{kmn}=\delta_{mi}\delta_{nj}-\delta_{mj}\delta_{ni}$. 
By making the correspondence $\vec{a}\rightarrow \rr-\RR_n$ and $\vec{b}\rightarrow\AA\left(\RR_n \ + \ (\rr-\RR_n) \ \lambda\right)$ in \eref{eq:Gn}, we have
\begin{eqnarray} \label{eq:Gn_bis}
		\nabla G_n(\rr) & = & 
		      \int_0^1 \! d\lambda \ [(\rr-\RR_n)\cdot\nabla] \AA\left(\RR_n \ + \ (\rr-\RR_n) \ \lambda\right)
		\ + \ \int_0^1 \! d\lambda \ [\AA\left(\RR_n \ + \ (\rr-\RR_n) \ \lambda\right)\cdot\nabla] (\rr-\RR_n)
		\\[3mm]&&
		\ + \ \int_0^1 \! d\lambda \  (\rr-\RR_n)\times[\nabla\times\AA\left(\RR_n \ + \ (\rr-\RR_n) \ \lambda\right)]
		\ + \ \int_0^1 \! d\lambda \  \AA\left(\RR_n \ + \ (\rr-\RR_n) \ \lambda\right)\times[\nabla\times(\rr-\RR_n)] \ .
\end{eqnarray}
If now we consider that 
\begin{equation}
	\left\{ \begin{array}{l}
  \nabla\times\AA\left(\RR_n \ + \ (\rr-\RR_n) \ \lambda\right) \ = \ \lambda  \ \BB\left(\RR_n \ + \ (\rr-\RR_n) \ \lambda\right) \\[3mm]
	\nabla\times(\rr-\RR_n) \ = \ 0 \\[3mm]
	[\AA\left(\RR_n \ + \ (\rr-\RR_n) \ \lambda\right)\cdot\nabla] (\rr-\RR_n) \ = \ \AA\left(\RR_n \ + \ (\rr-\RR_n) \ \lambda\right)
	\end{array}\right.
\end{equation}
we have
\begin{equation}
		\nabla G_n(\rr)  =  
		      \int_0^1 \! d\lambda \ [(\rr-\RR_n)\cdot\nabla] \AA\left(\RR_n  +  (\rr-\RR_n) \ \lambda\right)
		 +  \int_0^1 \! d\lambda \ \AA\left(\RR_n  +  (\rr-\RR_n) \ \lambda\right)
		 +  \int_0^1 \! d\lambda \ \lambda (\rr-\RR_n)\times \BB\left(\RR_n  +  (\rr-\RR_n) \ \lambda\right)
		\ .
\end{equation}
The second term on the right-hand side can be integrated by parts
\begin{equation}
	\left.\int_0^1 \! d\lambda \ \AA\left(\RR_n  +  (\rr-\RR_n) \ \lambda\right)
	\ = \ \lambda  \AA\left(\RR_n  +  (\rr-\RR_n) \ \lambda\right) \right|_{0}^{1}
	\ - \ \int_0^1 \! d\lambda \ \lambda \ \Frac{d}{d\lambda} \AA\left(\RR_n  +  (\rr-\RR_n) \ \lambda\right)
	\ = \ \AA(\rr) \ - \ \int_0^1 \! d\lambda \ \lambda \ [(\rr-\RR_n)\cdot\nabla] \AA\left(\RR_n  +  (\rr-\RR_n) \ \lambda\right) 
\end{equation}
and finally
\begin{equation}
		\nabla G_n(\rr)  \ = \  \AA(\rr) \ + \ \int_0^1 \! d\lambda \ \lambda (\rr-\RR_n)\times \BB\left(\RR_n  +  (\rr-\RR_n) \ \lambda\right) \ . 
\end{equation}
Since the basis is localized, we can consider $\rr\approx\RR_n$ and then the second term on the right-hand side of the above equation approximately vanishes and we finally obtain 
$\nabla G_n(\rr)  \approx  \AA(\rr)$, as we wanted to demonstrate.
%
\section{Properties of the vector potential for one-dimensional periodic systems} \label{app_vector1D}
In this appendix, we illustrate the properties of the vector potential of \eref{eq:peierls_general} with the gauge of \eref{eq:gauge1d}. 
Starting from 
\begin{equation}
\chi(\rr) \ = \ \Frac{1}{2} \ \rr\cdot\tt \ \rr\cdot\left(\tt\times\BB\right)
\ \ \ \rightarrow \ \ \ 
\nabla\chi(\rr) \ = \ \Frac{\rr\cdot\left(\tt\times\BB\right)}{2} \ \tt \ + \ \Frac{\rr\cdot\tt}{2} \ \tt\times\BB \ ,
\end{equation}
it follows that
\begin{equation}
\AA(\rr) \ = \ \Frac{1}{2} \ \BB\times\rr \ + \ \nabla\chi(\rr)
         \ = \ \Frac{1}{2} \ \BB\times\rr \ + \  \Frac{\rr\cdot\left(\tt\times\BB\right)}{2} \ \tt \ + \ \Frac{\rr\cdot\tt}{2} \ \tt\times\BB 
				\ = \ {\rm M}(\BB,\tt) \ \rr \ , 
\end{equation}
where 
\begin{equation}
{\rm M}(\BB,\tt) \ \equiv \ \Frac{1}{2} 
   \left(\begin{array}{ccc}
      -2 t_x t_z B_y \ + \ 2 t_x t_y B_z & t_x t_z B_x \ - \ t_y t_z B_y \ + \ (t_y^2-t_x^2-1) \ B_z & -t_x t_y B_x \ + \ (1+t_x^2-t_z^2) \ B_y \ + \ t_y t_z B_z \\[3mm]
			t_x t_z B_x \ - \ t_y t_z B_y \ + \ (1+t_y^2-t_x^2) \ B_z & 2 t_y t_z B_x \ - \ 2 t_x t_y B_z & (t_z^2-t_y^2-1) B_x \ + \ t_x t_y B_y \ - \ t_x t_z B_z \\[3mm] 
			-t_x t_y B_x \ + \ (t_x^2-t_z^2-1) B_y \ + \ t_y t_z B_z & (1+t_z^2-t_y^2) B_x \ + \ t_x t_y B_y \ - \ t_x t_z B_z & -2 t_y t_z B_x \ + \ 2 t_x t_z B_y 
   \end{array}\right)
\end{equation}
The eigenvalues of M are $\lambda=0,\ \pm i\BB\cdot\tt/2$. 
The eigenvalue $\lambda=0$ corresponds to the invariance of $\AA(\rr)$ along the direction identified by the unit vector $\tt$, which belongs to the kernel of the matrix M, i.e., $\AA(\rr=\alpha\tt)=0$ with $\alpha\in\mathbb{R}$. 
The other two complex eigenvalues correspond to closed field lines, which lie on a plane orthogonal to the vector $\hh=\BB-(\BB\cdot\tt)/2 \ \tt$. 
This can be verified by expressing a generic vector $\hh=\alpha \tt+\beta\BB+\gamma \tt\times\BB$ in terms of its components parallel to $\tt$, $\BB$, and their orthogonal direction $\tt\times\BB$. 
By imposing $\hh\cdot\AA=0$, we obtain $\gamma=0$ and $\alpha=-\beta\BB\cdot\tt/2$. 

\begin{figure}[!tb] 
	\centering
		\resizebox{8cm}{!}{\includegraphics{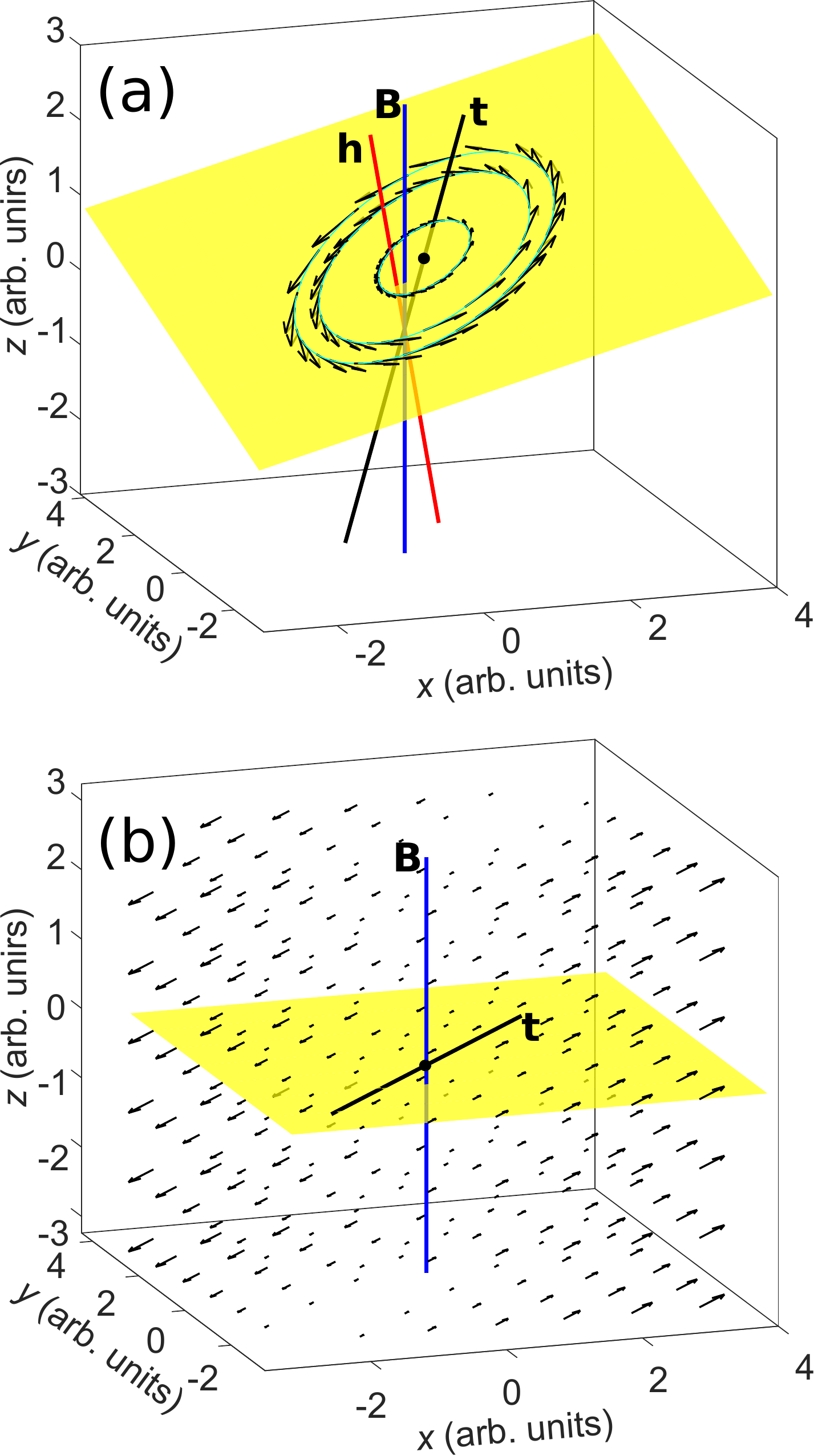}}
		\caption{\label{fig_a_vector_1d}
		   (a) Some $\AA(\rr)$ field vectors and lines in the case $\BB=(0,0,B)$ (blue line) and $\tt=(1,1,1)/\sqrt{3}$ (black line). 
			     The vector potential lies on planes (as the yellow one) that are orthogonal to $\hh=B(-1,-1,2)/3$ (red line), and turns around the intersection zero points along the direction identified by $\tt$, as indicated by a black dot in the example of the figure.
			 (b) $\AA(\rr)$ field vectors in the case of orthogonal $\BB=(0,0,B)$ and $\tt=(1,1,0)/\sqrt{2}$ (black line). 
					The vector potential is oriented along the direction identified by $\tt$ and therefore lies on planes orthogonal to $\BB$, as in the represented yellow plane.
				}
\end{figure}

\Fref{fig_a_vector_1d}(a) shows an example for $\BB=(0,0,B)$ along the $z$~direction (blue line) and periodicity along the direction $\tt=(1,1,1)/\sqrt{3}$ (black line). 
The vector potential $\AA$ vanishes along the direction $\tt$, and it lies on planes that are orthogonal to direction identified by the vector $\hh=B(-1,-1,2)/3$ (red line), such as the yellow plane in the figure. 
The field lines (cyan lines) are closed and the field vectors (black arrows) turn around the point (in black) at the intersection between the plane and the direction identified by $\tt$. 

If the magnetic field and the periodicity direction are orthogonal, i.e., $\BB\cdot\tt=0$, then, after some algebra, we obtain $\AA(\rr)=\rr\cdot(\tt\times\BB)\tt$. 
Therefore, the vector potential is oriented along $\tt $ and has a length given by the projection of $\rr$ along $\tt\times\BB$. 
In particular, if $\BB=(0,0,B)$ and $\tt=(1,0,0)$, as mentioned in \sref{sec:explicit_peierls} and as in one of the cases considered for the carbon nanotubes in \sref{subsec:cnts}, we obtain the first Landau gauge $\AA=(-By,0,0)$. 
Note that when $\BB\cdot\tt=0$, the three eigenvalues of M are identically 0. 
This confirms that the field lines, which are parallel lines oriented as $\tt$, do not close on themselves.
\Fref{fig_a_vector_1d}(b) shows an example for $\BB=(0,0,B)$ along the $z$~direction (blue line) and along the direction $\tt=(1,1,0)/\sqrt{2}$ (black line).

\section{From two-dimensional to one-dimensional systems} \label{app:2to1}
In this appendix, we show how the Peierls phase factors of \eref{eq:peierls_phase_1d} for periodic 1D systems can be obtained from that of \eref{eq:recipe_2d_general} for periodic 2D systems.
Starting from the 2D system, we can pass to the 1D system by considering $\TT_1=\TT$, $m_1=m$, $n_1=n$, and $m_2=n_2=0$. 
We can also arbitrarily set $\TT_2\parallelslant\BB$, so that $\TT_1$, $\TT_2$ and $\BB$ are coplanar and we can make use of \eref{eq:coplanar_B} without conditions on the magnetic field value. 
It follows that
\begin{equation} \label{eq:1from2}
 \varphi_{i,m;j,n} \ \equiv \ \varphi_{i,m,0;j,n,0} \ = \ \Frac{\pi}{\Phi_0} \BB\cdot\left[ \dd_i\times\dd_j + (\dd_i+\dd_j) \times (n-m)\TT + \chi_{j00} - \chi_{i00}\right] \ ,
\end{equation}
where we reintroduce the gauge $\{\chi_{i00}\}$, which previously was arbitrarily set to 0 in \eref{eq:chi_chi}.
We now demonstrate that \eref{eq:1from2} is equivalent to \eref{eq:peierls_phase_1d}. 
To do this, we consider a periodic 1D system or subsystem where the basis states are identified by the positions $\RR_{im}=\dd_i+m\TT$. 
From \eref{eq:peierls_phase_1d}, we have
\begin{eqnarray} 
  \varphi_{i,m;j,n} & = & \Frac{\pi}{\Phi_0} \ \BB \cdot \left[ (\dd_i+m\TT) \times (\dd_j+n\TT) 
																                    \ + \ (\dd_j+n\TT)\cdot\Frac{\TT}{|\TT|} \ \ (\dd_j+n\TT) \times \Frac{\TT}{|\TT|} \ 
							                    	                \ - \ (\dd_i+m\TT)\cdot\Frac{\TT}{|\TT|} \ (\dd_i+m\TT) \times \Frac{\TT}{|\TT|} \
																										\right] \nonumber
			 		\\[4mm]   & = & \Frac{\pi}{\Phi_0} \ \BB \cdot \left[ \dd_i\times\dd_j \ + \ (\dd_i+\dd_j) \times (n-m) \TT 
																                    \ + \ \dd_j\cdot\Frac{\TT}{|\TT|} \ \dd_j\times\Frac{\TT}{|\TT|} \ 
							                    	                \ - \ \dd_i\cdot\Frac{\TT}{|\TT|} \ \dd_i\times\Frac{\TT}{|\TT|} \
																										\right] \ . \label{eq:peierls_phase_1d_special}																										
\end{eqnarray}
\Eref{eq:1from2} and \eref{eq:peierls_phase_1d_special} are exactly the same provided we make the gauge choice $\chi_{i00}=\left(\dd_i\cdot\TT/|\TT|\right) \ \dd_i\times\TT/|\TT|$.

%

\end{document}